\documentclass[journal]{IEEEtran}

\usepackage{amsmath}
\usepackage{amssymb}
\usepackage{amsfonts}
\usepackage{graphicx}
\usepackage{epsfig}
\usepackage{subfigure}
\usepackage{psfrag}
\usepackage{cite}
\usepackage{latexsym}
\usepackage{url}
\usepackage{color}
\usepackage{multirow}
\usepackage{mathtools}
\usepackage{bm}
\usepackage{booktabs}
\usepackage{algorithm}
\usepackage{algpseudocode}

\usepackage{verbatim}
\usepackage{indentfirst}
\usepackage{hyperref}

\PassOptionsToPackage{bookmarks={false}}{hyperref}

\IEEEoverridecommandlockouts

\newtheorem{theorem}{\underline{Theorem}}[section]
\newtheorem{lemma}{\underline{Lemma}}[section]

\newtheorem{remark}{\underline{Remark}}[section]
\newcommand{\mv}[1]{\mbox{\boldmath{$ #1 $}}}

\begin{document}
\title{Rethinking Resource Management in Edge Learning: A Joint Pre-training and Fine-tuning Design Paradigm}
\author{
Zhonghao Lyu, \emph{Graduate Student Member, IEEE,} Yuchen Li, Guangxu Zhu, \emph{Member, IEEE,} \\ Jie Xu, \emph{Senior Member, IEEE,}  H. Vincent Poor, \emph{Life Fellow, IEEE,} and Shuguang Cui, \emph{Fellow, IEEE}\\
\thanks{Z. Lyu is with the Future Network of Intelligence Institute (FNii) and the School of Science and Engineering (SSE), The Chinese University of Hong Kong (Shenzhen), Shenzhen, China (e-mail: zhonghaolyu@link.cuhk.edu.cn). }
\thanks{Y. Li is with the Department of Computer Science and Engineering, Shanghai Jiao Tong University, Shanghai, China, and also with Baidu, Inc., Beijing, China (e-mail: yuchenli@sjtu.edu.cn).}
\thanks{J. Xu and S. Cui are with the SSE and FNii, The Chinese University of Hong Kong (Shenzhen), Shenzhen, China (e-mail: xujie@cuhk.edu.cn, shuguangcui@cuhk.edu.cn). J. Xu is the corresponding author.}
\thanks{
G. Zhu is with Shenzhen Research Institute of Big Data, Shenzhen, China  (e-mail: gxzhu@sribd.cn).}
\thanks{H. V. Poor is with the Department of Electrical and Computer
Engineering, Princeton University, Princeton, USA (e-mail:
poor@princeton.edu).}
}
\maketitle

\begin{abstract}
In some applications, edge learning is experiencing a shift in focusing from conventional learning from scratch to new two-stage learning unifying pre-training and task-specific fine-tuning. This paper considers the problem of joint communication and computation resource management in a two-stage edge learning system. In this system, model pre-training is first conducted at an edge server via centralized learning on local pre-stored general data, and then task-specific fine-tuning is performed at edge devices based on the pre-trained model via federated edge learning. For the two-stage learning model, we first analyze the convergence behavior (in terms of the average squared gradient norm bound), which characterizes the impacts of various system parameters such as the number of learning rounds and batch sizes in the two stages on the convergence rate. Based on our analytical results, we then propose a joint communication and computation resource management design to minimize an average squared gradient norm bound, subject to constraints on the transmit power, overall system energy consumption, and training delay. The decision variables include the number of learning rounds, batch sizes, clock frequencies, and transmit power control for both pre-training and fine-tuning stages. Finally, numerical results are provided to evaluate the effectiveness of our proposed design. It is shown that the proposed  joint resource management over the pre-training and fine-tuning stages well balances the system performance trade-off among the training accuracy, delay, and energy consumption. The proposed design is also shown to effectively leverage the inherent trade-off between pre-training and fine-tuning, which arises from the differences in data distribution between pre-stored general data versus real-time task-specific data, thus efficiently optimizing overall system performance.
\end{abstract}

\section{Introduction}
In recent decades, artificial intelligence (AI) technologies have achieved considerable success  in various areas, such as natural language processing (NLP) \cite{DWOtter} and computer vision (CV) \cite{YGuo}. In particular, the recent breakthroughs in Large Language Models (LLMs) (such as OpenAI's ChatGPT \cite{OpenAI}) have empowered machines with human-like content comprehension and generation capability. Within this context,  AI is playing a key role in creating a more intelligent and collaborative information infrastructure.

At the same time, advances in wireless communications technologies have accelerated the ongoing convergence of AI and wireless networks, forming the foundation of today's emerging  AI-native networks \cite{GZhuDLiu,KZhao,GZhu2023}. Conventionally, the training and inference of AI algorithms rely on the cloud AI paradigm, where massive amounts of data are offloaded to the cloud for centralized processing. However, facing the challenges of massive datasets and limited network connectivity as well as ultra-low latency requirements, cloud AI may fail to support emerging applications such as industrial control, autonomous vehicles, driver-assist systems, and virtual reality (VR). To tackle such issues, edge AI has become an alternative solution to deliver AI services at the network edge, by bringing computing power
and AI functionality close to data via the exploitation of mobile edge computing (MEC) techniques \cite{ZZhou2019}.

Particularly, the efficient learning of AI models at the network edge can be generally divided into two categories, namely, centralized and distributed edge learning. In centralized edge learning, raw data is required to be uploaded to edge servers for further processing, which may cause extremely heavy transmission burdens. By contrast, in distributed edge learning, separate edge servers and devices can fully utilize the rich distributed computation and data resources for collaboratively training machine learning models. Among various distributed edge learning approaches, federated edge learning (FEEL) has emerged as an important technology due in part to its ability to help protect privacy \cite{XCaoMagazine,LYou}. In FEEL, the joint design of communication and computation is a key issue to combat  wireless channel fading and efficiently utilize shared resources. 

A number of prior studies have investigated centralized edge learning and FEEL separately. For centralized edge learning, data-importance aware resource management schemes (such as user scheduling and power control) \cite{DLiu2021} have been widely adopted in existing works. For FEEL, existing works have studied the performance analysis and resource management \cite{YYang,YJiang,YShi2023,HZhao,ZYang,MSAlAbiad,SLuo, XHan,ZChen,YSun,YLi2023,YLi2024,XMo2021}, as well as other novel enabling technologies (such as over-the-air computation (AirComp) \cite{JDu2023}) to improve learning and energy efficiency. For example, in order to enhance the \emph{learning efficiency} (e.g., learning performance and delay), the authors in \cite{YYang,YJiang,YShi2023,HZhao} analyzed the convergence behavior (in terms of the optimality gap) by considering the effects of  model (gradient) quantization, sparsification, client selection, and communication noise, respectively. Based on the convergence analysis, they conducted  joint resource  management for minimizing the  training latency  or enhancing the learning performance via minimizing the optimality gap. On the other hand, to improve the \emph{energy efficiency}, the authors in \cite{ZYang,MSAlAbiad,SLuo, XHan,ZChen,YSun,YLi2023,YLi2024,XMo2021} optimized the joint allocation of computation (such as central processing unit (CPU)-cycle frequencies) and communication (such as transmit power, bandwidth, and client scheduling) resources in centralized \cite{ZYang, XHan,ZChen,YSun,YLi2023,YLi2024,XMo2021}, decentralized \cite{MSAlAbiad}, and hierarchical FEEL \cite{SLuo}, respectively. Moreover, other works have further exploited model/gradient compression \cite{LLi2021} and data selection \cite{ZChen2023} techniques to enhance energy efficiency by properly reducing  the total communication and computation workloads while preserving the learning performance.

Note that  the above works  considered training AI models from scratch with centralized/distributed edge learning. Recently, however, multi-stage learning paradigms have emerged for the efficient training of AI models at the network edge, where the learning process can be generally divided into  two stages of \emph{pre-training} and \emph{task-specific fine-tuning} \cite{LBariah, JWang, YShen}. This new paradigm  may greatly alleviate the challenge of significant time and energy consumption caused by training from  scratch, especially when the size of AI models increases dramatically\footnote{By taking GPT-3 as an example, training it roughly takes 34 days on 1024 A100 graphics processing units (GPUs) \cite{DNarayanan},  requires 1,287 MWh energy consumption, and emits 552 tonnes C${\rm O}_2$eq \cite{ASLuccioni}.}.  Specifically, during the \emph{pre-training} stage, the model is extensively trained on vast amounts of data to enlarge the generalization ability of the model via acquiring inherent statistical semantic representations. Such extensive training is clearly prohibitive on edge devices, but it could be conducted on local  pre-stored data in a centralized learning manner. For the \emph{task-specific fine-tuning} stage, the pre-trained AI model can be further fine-tuned on edge devices using task-specific data to better align with the target task. In particular,
FEEL provides a well acknowledged
solution for task-specific fine-tuning at the network edge to fully leverage the distributed computation power and the timely task-specific data resources \cite{AHilmkil, SSavazzi,MXu}. 



In view of the above issues, it is of great interest to rethink edge learning design by considering its entire life-cycle of implementation, which brings pre-training and task-specific fine-tuning together. The integration of pre-training and task-specific fine-tuning introduces various new technical challenges. On the one hand, the pre-trained model is normally adopted as the initial model for fine-tuning. Therefore, the learning performance (e.g., the convergence behavior) depends on both stages. It is thus   important to analyze the learning performance to characterize the effect of both stages. However, the differences in data distribution and training procedures in the two stages make such an analysis difficult. On the other hand, both pre-training and fine-tuning may cause considerable training delay and energy consumption. There generally exists a delay (and also energy) trade-off over the two stages in achieving a certain overall learning performance. However, the trade-off relationship is complicated in general due to the heterogeneity of edge servers and devices in terms of communication and computation resources. Therefore,  it is also important to perform joint resource management to well balance such trade-offs. These issues have not been well investigated in the existing literature, which motivates our work in this paper. 

In particular, in this paper we propose a new two-stage edge learning framework, which considers the entire life-cycle of edge learning unifying pre-training and task-specific fine-tuning. In this framework, the model pre-training stage is conducted by centralized learning with local pre-stored data. After model pre-training, task-specific fine-tuning is performed based on FEEL in a distributed manner for further performance enhancement. For this system, we analyze the convergence behavior, energy consumption, and delay. Based on the
analytical results, we further design a joint resource management approach for learning performance enhancement. The main contributions are elaborated as follows.
\begin{itemize}
	\item First,  under the considered two-stage edge learning system unifying model pre-training and task-specific fine-tuning, we analyze the convergence behavior of the proposed framework in terms of an average squared gradient norm bound, which captures the influence of various important system parameters including the number of learning rounds, batch sizes, and data distributions across the pre-training and fine-tuning stages.
	\item  Next, building on the convergence analysis, we formulate an average squared gradient norm bound minimization problem, by jointly optimizing the communication and computation resources in the  pre-training and fine-tuning stages (including the number of learning rounds, batch sizes, clock frequencies, and transmit power),
	subject to a set of constraints on the transmit power, overall energy consumption, and training delay. Due to the close coupling of the computation and communication variables, the problem is highly non-convex
	and thus non-trivial to solve. To tackle this problem, we propose an efficient algorithm  that can obtain a high-quality sub-optimal solution by using two-dimensional search and successive convex approximation (SCA) techniques.
	\item Finally, extensive numerical results are provided to evaluate the performance of the proposed design under the joint pre-training and fine-tuning framework. It is shown that the proposed design achieves lower average gradient norm bound (or loss values) than  benchmark schemes under the same system delay and energy consumption. This is due to the fact that the proposed design properly exploits the inherent trade-off between pre-training and fine-tuning arising from the variations in data distributions over the two stages, thus efficiently optimizing the performance of the considered two-stage edge learning system.
\end{itemize}

\begin{figure*}[h]
	\centering
	 \epsfxsize=1\linewidth
		\includegraphics[width=14cm]{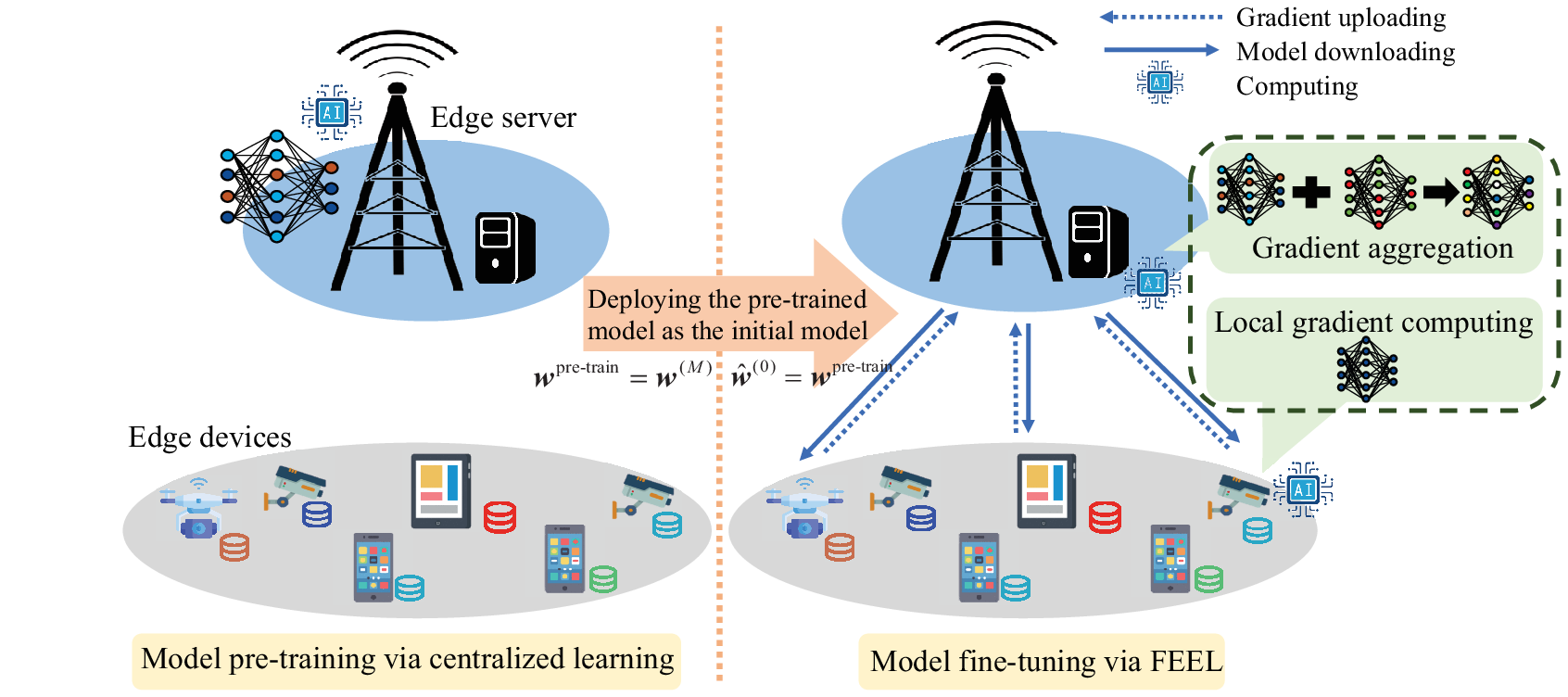}
	\caption{\label{model}Illustration of the considered two-stage edge learning system with model pre-training and task-specific fine-tuning.}
	\vspace{-10pt}
\end{figure*}

The remainder of this paper is organized as follows. Section II introduces the  two-stage edge learning system model. Section III analyzes the convergence behavior of the system. Section IV formulates the average squared gradient norm bound minimization problem and provides an efficient solution. Section V presents numerical results to show the effectiveness  of the proposed design. Section VI concludes this paper.


\section{System Model}
We consider a two-stage edge learning system consisting of an edge server and $K > 1$ edge devices, which  cooperatively learn a deep learning (DL) model as shown in Fig. \ref{model}.  It is assumed that the edge server has pre-stored general data that can be used for pre-training, and edge devices have most recent task-specific data. The entire edge learning process is divided into two stages. First, the edge server learns a pre-trained DL model via centralized learning on the pre-stored general data. Next, the pre-trained model is deployed at edge devices for task-specific fine-tuning via FEEL. Specifically, FEEL is implemented in an iterative manner. In each iteration, the edge devices learn local models with their local data, then the edge server aggregates local models to obtain a global model, and finally, the edge server broadcasts the obtained global model to edge devices for their local update in the next round. In the following, we introduce the above two stages, respectively.

\subsection{Pre-training via Centralized Learning}
In order to speed up the learning process on a specific target task, it is highly desired to perform model pre-training beforehand to obtain a high-quality initial model with the perception of general underlying features.

Suppose that the pre-trained model is represented by the parameter vector $\mv{w} \in \mathbb{R}^q$ with $q$ denoting the model size. Denote ${\mathcal D}$ as the local training data pre-stored at the edge server with distribution $\mathcal{P}$. Denote the size of ${\mathcal D}$ as $|{\mathcal D}| = D \ge 1$, where $|{\mathcal D}|$ is the cardinality of ${\mathcal D}$. We employ the stochastic gradient descent (SGD) to train the model for $M$ rounds with ${\cal M}\triangleq\{0,\cdots,M-1\}$. For each round $m \in {\cal M}$, let $l(\mv{w}^{(m)}, d_{i})$ denote the loss function of model $\mv{w}^{(m)}$ on sample $d_{i} \in {\mathcal D}$. Accordingly, the total expected loss over ${\mathcal D}$ is
\begin{align}
	{L}(\mv{w}^{(m)},{\cal D}) = \mathbb{E}_{{d_{i}} \sim \mathcal{P}} \left[{l(\mv{w}^{(m)}, d_{i})} \right].
\end{align}

To train the model, we randomly select a mini-batch of data $\tilde{\mathcal D}^{(m)}$ with $|\tilde{\mathcal D}^{(m)}| = \tilde{D}^{(m)} $ for each round $m$. The local loss estimation over sampled data $\tilde{\mathcal D}^{(m)}$ is
\begin{align}\label{local loss estimate}
{\tilde{L}}(\mv{w}^{(m)},\tilde{\mathcal D}^{(m)})= \frac{1}{\tilde{D}^{(m)}}\sum\nolimits_{{\tilde{d}^{(m)}_i} \in {\tilde{\mathcal D}^{(m)}}} {l(\mv{w}^{(m)}, {\tilde{d}^{(m)}_i})}.
\end{align}
Based on \eqref{local loss estimate}, we update the model for round $m+1$ as
\begin{align}\label{parameter update pre-train}
	{{\mv{w}}^{(m+1)}} = {{\mv{w}}^{(m)}}  - \gamma  \times {\nabla _{{{\mv{w}}^{(m)}}}}{\tilde{L}}(\mv{w}^{(m)},\tilde{\mathcal D}^{(m)}),
\end{align}
where $\gamma$ denotes the learning rate.

The whole pre-training process iterates for $M$ rounds, and then we obtain the ultimate pre-trained model $\mv{w}^{\rm {pre-train}}=\mv{w}^{(M)}$. 

\subsection{Task-specific Fine-tuning via FEEL}

Next, based on the pre-trained model $\mv{w}^{\rm {pre-train}}$, we proceed with the task-specific fine-tuning via FEEL to align with the target task. In FEEL, the set $\cal K$ of edge devices cooperatively train a shared machine learning model with the coordination of the edge server, as elaborated in the following.

Let ${\mathcal{B}}_k$ denote the stored data of each edge device $k \in \mathcal K$ for task-specific fine-tuning with $|{\mathcal{B}}_k|={B}_k$. Assume that ${\mathcal{B}}_k$ shares the same distribution of $\hat{\mathcal{P}}$ for all devices $k \in \mathcal K$. Note that the distribution $\hat{\mathcal{P}}$ is generally different from $\mathcal{P}$ for the pre-stored data at the edge server. Moreover, denote ${l}(\hat{\mv{w}}, {b}_{k,j})$ as the sample-wise loss function of model $\hat{\mv{w}}$ (with the same size of $\mv{w}$) during FEEL on sample ${b}_{k,j} \in {\mathcal{B}}_k$. Then the local
expected loss function of edge device $k$  on ${\mathcal{B}}_k$ is
\begin{align}
	{L_k}(\hat{\mv{w}},{\mathcal{B}}_k) = \mathbb{E}_{{{b}_{k,j}} \sim \hat{\mathcal{P}}} \left[l(\hat{\mv{w}}, {b}_{k,j}) \right] .
\end{align}
Accordingly, the global loss function on the whole dataset ${{\mathcal B}}\triangleq \cup_{k=1}^K {{\mathcal B}}_k$ of all $K$  devices is given by
\begin{align} \label{global loss}
{L}(\hat{\mv{w}}, {{\mathcal B}}) =  \frac{1}{K}\sum\limits_{k \in {\cal K}} {L_k}(\hat{\mv{w}},{\mathcal{B}}_k) .
\end{align}

In general, the objective of the fine-tuning process is to minimize the global loss function  in \eqref{global loss} to obtain the optimal model parameter as
\begin{align} \label{optimal w}
	\hat{\mv w}^{*} = \arg \mathop {\min }\limits_{\hat {\mv w}} {L}(\hat{\mv{w}}, {{\mathcal B}}).
\end{align}

Next, we employ the FEEL to solve \eqref{optimal w} in a distributed manner. It is worth noticing that we consider federated stochastic gradient descent (FedSGD) algorithm \cite{HBMcMahan2017} for simplicity in this paper, where a single local iteration gradient update is considered in each learning round. However, the proposed  framework could be extended to a more general setup with multiple local iterations  in each learning round (e.g., federated averaging (FedAvg) \cite{XLi2019}). Specifically, let $ N$ denote the number of learning rounds, and define ${\cal N}\triangleq\{0,\cdots,N-1\}$. At each round $n \in {\cal N}$, the edge server first broadcasts the global model $\hat{{\mv w}}^{(n)} $ to all edge servers, where the initial global model at round $n=0$  is the pre-trained model $\hat{{\mv w}}^{(0)}=\mv{w}^{\rm {pre-train}}$. Next,  each device $k \in \mathcal{K}$ randomly selects a mini-batch of samples $\tilde{{{\mathcal B}}}^{(n)}_{k}$ with batch size $\tilde{{{ B}}}^{(n)}_{k}=|\tilde{{{\mathcal B}}}^{(n)}_{k}|$. Based on $\tilde{{{\mathcal B}}}^{(n)}_{k}$ and $\hat{{\mv w}}^{(n)}$, the device computes the
local gradient as
\begin{align} 
	\nabla \tilde{{ L}}^{(n)}_{k} (\hat{{\mv w}}^{(n)}, \tilde{{{\mathcal B}}}^{(n)}_{k})= \frac{1}{\tilde{{{ B}}}^{(n)}_{k}} \sum\nolimits_{\tilde{{b}}^{(n)}_{k,j} \in \tilde{{{\mathcal B}}}^{(n)}_{k}} \nabla {{l}(\hat{{\mv w}}^{(n)}, \tilde{{b}}^{(n)}_{k,j})}. 
\end{align}

Then the edge devices upload the obtained local gradients to the edge server, and the server aggregates them to obtain the overall gradient as 
\begin{align} 
	\nabla \bar{L}^{(n)} (\hat{{\mv w}}^{(n)}, \{\tilde{{{\mathcal B}}}^{(n)}_{k}\}) = \sum\limits_{k \in {{\mathcal K}}} \frac{\tilde{{B}}^{(n)}_{k}}{\tilde{{B}}^{(n)}_{\rm tot}} \nabla \tilde{L}^{(n)}_{k} (\hat{{\mv w}}^{(n)}, \tilde{{{\mathcal B}}}^{(n)}_{k}), 
\end{align}
where $\tilde{{B}}^{(n)}_{\rm tot}=\sum\nolimits_{k \in {{\mathcal K}}} \tilde{{B}}^{(n)}_{k}$ denotes the total size of training data over the  $K$ devices. Then, with learning rate $\hat{{\gamma}}$, the global model at the edge server is uploaded by
\begin{align} \label{FEEL parameter update}
	\hat{{\mv w}}^{(n+1)} = \hat{{\mv w}}^{(n)} - \hat{{\gamma}}\nabla \bar{L}^{(n)} (\hat{{\mv w}}^{(n)}, \{\tilde{{\mathcal B}}^{(n)}_{k}\}).
\end{align}

Finally, the edge server sends the updated global  model back to all devices, and starts the next learning round. The above procedure iterates until the maximum number of learning rounds $N$ is met.

\subsection{Communication Model}
We first present the communication model for gradient uploading, in which edge devices upload the gradients via frequency-division multiple access (FDMA). Let ${\mv x}^{(n)}_k$ denote the transmitted signal by the $k$-th device, and we assume each element ${x}^{(n)}_{k,i}$ of ${\mv x}^{(n)}_k$ has normalized power, i.e., $\mathbb{E}[|{x}^{(n)}_{k,i}|^2]=1$, without
loss of generality. Then denote $p^{(n)}_{k}$ as the transmit power of device $k$ at round $n$, which should satisfy the average and maximum power constraints as 
\begin{align} \label{average power constraints}
	\frac{1}{N}\sum\limits_{n \in \mathcal{N}} {p^{(n)}_{k} \le P_k^{\rm ave}} ,\forall k \in \cal K,\\
	0\le {p^{(n)}_{k} \le P_k^{\rm max}} ,\forall n \in {\cal N}, k \in \cal K,
\end{align}
where $P_k^{\rm ave}$ and $P_k^{\rm max}$ denote the average and maximum power budgets of device $k$ over the whole FEEL period. Denote $g^{(n),\rm u}_{k}$ as the uplink channel power gain from edge device $k$ to the edge server at learning round $n$. For ease of exposition, the wireless channels are assumed to be unchanged over each learning round. Then the achievable rate by edge device $k$ (in bits-per-second (bps)) at learning round $n$ is
\begin{align} 
	r^{(n), \rm u}_{k}\left(p^{(n)}_{k}\right)=W_k^{\rm u}\log_2 \left(1+\frac{g^{(n),\rm u}_{k} p^{(n)}_{k}}{ W_k^{\rm u}N_0^{(n)}} \right),
\end{align}
where $W_k^{\rm u}$ is the allocated bandwidth for device $k$, $N_0^{(n)}$ is the power spectrum density (PSD) of the additive white Gaussian noise (AWGN) at the receiver of the edge server  at round $n$.

Next, we present the communication modeling for model downloading. Denote $g^{(n), \rm d}_{k}$ as the channel power gain from the edge server to edge device $k$ at learning round $n$. The downlink communication rate for device $k$ at learning round $n$ is
\begin{align} 
	{r}^{(n),\rm d}_{k}=W^{\rm d}\log_2\left(1+\frac{g^{(n),\rm d}_{k} \tilde{P}}{ W^{\rm d} \tilde{N}_{0,k}^{(n)} }    \right),
\end{align}
where $W^{\rm d}$ denotes the   bandwidth for downlink transmission,  $\tilde{P}$ denotes the transmit power of the edge server, and $\tilde{N}_{0,k}^{(n)}$ is the PSD of AWGN at the receiver of device $k$.

\subsection{Training Delay and Energy Consumption Analysis}

\subsubsection{Training Delay Analysis}First, we analyze the overall delay of the proposed system, which consists of the delay  for both pre-training and fine-tuning (including local training, gradient uploading, and model downloading) stages. 

First, we consider the delay for the pre-training stage. Denote $N_{\rm FLOP}$, $f^{(m)}$, and $c$ as the number of floating point operations (FLOPs) for gradient computation for processing one data sample, the clock frequency for round $m$, and the number of FLOPs per cycle of the processor (such as CPU and GPU). Then the training delay for the pre-training stage with $M$ learning rounds is 
\begin{align}\label{delay_pretrain}
	\tau(M,\{\tilde{D}^{(m)}\},\{f^{(m)}\})= \sum\limits_{m = 0}^{M-1}  \frac{\tilde{D}^{(m)} N_{\rm FLOP} }{f^{(m)} c}.
\end{align}

Next, we derive  the delay for the FEEL fine-tuning stage, including that for local training, gradient uploading, and model downloading. Similarly, denote ${\hat f}_k^{(n)}$ and ${\hat c}_k$ as the clock frequency and the number of FLOPs per cycle of the processor at device $k$. Then the computation delay  for local training of device $k$ at each round $n$ is 
\begin{align}\label{delay_comp_finetune}
\hat{\tau}^{(n),\rm {t}}_{k}(\tilde{{B}}^{(n)}_{k})=\frac{\tilde{{B}}^{(n)}_{k} N_{\rm FLOP}}  {{\hat f}_k^{(n)} {\hat c}_k}. 
\end{align}
Next, the gradient uploading duration for device $k$ at round $n$ is
$\hat{\tau} ^{(n), \rm {u}}_{k} (p^{(n)}_{k})= \beta/r^{(n),\rm u}_{k}(p^{(n)}_{k})$, where $\beta$ denotes the gradient (model) size (in bit). Accordingly, the model downloading duration is $\hat{\tau} ^{(n),\rm d}_{k} = \beta/{r}^{(n),\rm d}_{k}$.

By combining the above components, the overall training delay of the  two-stage edge learning system is
\begin{align}\label{delay_overall}
	&\tilde{\tau}(M,N,\{\tilde{D}^{(m)}\},\{f^{(m)}\},\{\tilde{{B}}^{(n)}_{k}\},\{{\hat f}_k^{(n)}\},\{p^{(n)}_{k}\}) \nonumber \\
	&= \tau(M,\{\tilde{D}^{(m)}\},\{f^{(m)}\})+ \sum\limits_{n = 0}^{N-1} \mathop {\max }\limits_{k \in {\cal K}} \left(\hat{\tau} ^{(n),\rm d}_{k} \right. \nonumber \\
	& \qquad \qquad \left. +  \hat{\tau}^{(n),\rm {t}}_{k}(\tilde{B}^{(n)}_{k}) + \hat{\tau} ^{(n), \rm {u}}_{k} (p^{(n)}_{k})\right) \nonumber\\
	&= \sum\limits_{m = 0}^{M-1}  \frac{\tilde{D}^{(m)} N_{\rm FLOP} }{f^{(m)}c} + \sum\limits_{n = 0}^{N-1} \mathop {\max }\limits_{k \in {\cal K}} \left(\frac{\beta}{{r}^{(n),\rm d}_{k}} \right. \nonumber \\
	&\qquad \qquad \left. +\frac{\tilde{B}^{(n)}_{k} N_{\rm FLOP}} {\hat{f}_k^{(n)} \hat{c}_k}+\frac{{\beta}}{r^{(n),\rm u}_{k}\left(p^{(n)}_{k}\right)}\right).
\end{align}

\subsubsection{Energy Consumption Analysis}
Furthermore,  we analyze the energy consumption of the proposed system. In the following, we present the energy consumption for pre-training and  fine-tuning, respectively. 

First, we consider the energy consumption in the pre-training stage. Specifically, for a complementary metal-oxide semiconductor (CMOS) circuit, the power consumption of the processor at the edge server with clock frequency $f$ is modeled as $\phi {f}^3$ \cite{QZeng2021}, where $\phi$ is the power coefficient depending on the chip architecture. Then the energy consumption for the pre-training stage\footnote{It is worth noticing that we focus on dynamic energy consumption (relevant to computational frequencies, transmit power, etc.) and ignore  static parts (such as energy of leakage currents and cooling) in the analysis of computation and communication energy consumption. Our proposed design can be readily  extended to the case considering the static energy consumption.} is
\begin{align}
	E(M,\{\tilde{D}^{(m)}\},\{f^{(m)}\})=\sum\limits_{m = 0}^{M-1}  \eta \frac{\tilde{D}^{(m)} N_{\rm FLOP} }{c}\phi {f^{(m)}}^2,
\end{align}
where $\eta$ denotes the power usage effectiveness (PUE) of the edge server.

Next, we focus on the energy consumption of task-specific fine-tuning via FEEL, which includes that for local training, gradient uploading, and model downloading, respectively. First, at each learning round $n$, the energy consumption of  local gradient computing for local training at device $k$  is 
\begin{align}
{\hat E}^{(n),\rm {t}}_{k}(\tilde{{B}}^{(n)}_{k},{\hat f}_k^{(n)})=\hat{\eta}_k \frac{ {\tilde{{B}}^{(n)}_{k} N_{\rm FLOP} } } {{\hat{c}_k}}\hat{\phi}_k {{\hat f}_k}^{(n)2},
\end{align}
where $\hat{\eta}_k$ is the PUE of device $k$, $\hat{\phi}_k$ is the power coefficient depending on its chip architecture. In addition, the energy consumption of gradient uploading for device $k$ is 
\begin{align}
E ^{(n), \rm u}_{k} (p^{(n)}_{k})= \frac{p^{(n)}_{k}{\beta}}{r^{(n),\rm u}_{k}(p^{(n)}_{k})},
\end{align}
and the energy consumption of edge server  for model downloading at round $n$ is 
\begin{align}
\hat{E} ^{(n),\rm d} = \tilde{P} \times \mathop {\max }\limits_{k \in {\cal K}} (\frac{\beta}{{r}^{(n),\rm d}_{k}}).
\end{align}
By combing the above components, the overall energy consumption of the proposed system over both pre-training and fine-tuning stages is
\begin{align}\label{energy_overall}
	&{\tilde E}\left( M,N,\{\tilde{D}^{(m)}\},\{f^{(m)}\},\{\tilde{{B}}^{(n)}_{k}\},\{{\hat f}_k^{(n)}\},\{p^{(n)}_{k}\} \right) \nonumber \\
	&= E(M,\{\tilde{D}^{(m)}\},\{f^{(m)}\})+ \sum\limits_{n = 0}^{N-1} \left( \hat{E} ^{(n),\rm d} + \right. \nonumber \\
	& \qquad \qquad \left. \sum\limits_{k = 1}^K \left({\hat E}^{(n),\rm {t}}_{k}\big(\tilde{{B}}^{(n)}_{k},{\hat f}_k^{(n)}\big)+E ^{(n), \rm u}_{k} (p^{(n)}_{k}) \right) \right)\nonumber\\
	&=\sum\limits_{m = 0}^{M-1}  \eta \frac{\tilde{D}^{(m)} N_{\rm FLOP} }{c}\phi {f^{(m)}}^2 + \sum\limits_{n = 0}^{N-1} \left( {\tilde P} \times \mathop {\max }\limits_{k \in {\cal K}} (\frac {\beta}{{r}^{(n),\rm d}_{k}}) +\right. \nonumber \\
	& \qquad \left.  \sum\limits_{k = 1}^K \left(   \hat{\eta}_k \frac{ {\tilde{B}^{(n)}_{k} N_{\rm FLOP} } \hat{\phi}_k {{\hat f}_k^{(n)2}}}    {\hat{c}_k} + \frac{p^{(n)}_{k}{\beta}}{r^{(n),\rm u}_{k}(p^{(n)}_{k})} \right)\right).
\end{align}

\section{Convergence Analysis}
\subsection{Assumptions and Definitions on Learning Models}
To facilitate the convergence analysis of the proposed two-stage edge learning framework, we make the following assumptions and definitions on the loss function and gradient estimate, as commonly adopted in existing works \cite{LBottou,YSun,NTripuraneni}. 

{\bf Assumption 1} (Bounded loss function): For any ${\hat {\mv w}} \in \mathbb{R}^q$, the global loss function is lower bounded by ${\hat L}_{\rm inf}$, i.e., $L(\hat {\mv w}, {\cal B}) \ge {\hat L}_{\rm inf}$.

{\bf Assumption 2} (Smoothness): The gradient ${\nabla}{L}(\mv{w}, \cal D)$ ($\nabla {L}({\hat{\mv w}},\cal B)$) is $\rho/\hat{\rho}$-Lipschitz continuous with respect to (w.r.t.) network parameters, where  $\rho$ and $\hat{\rho}$ are parameters related to pre-training and fine-tuning stages, respectively. In other words,  for any ${\mv w},{\mv v}, \hat{\mv w},\hat{\mv v} \in \mathbb{R}^q $ in pre-training and fine-tuning stages, it follows that
\begin{align}
	\| {\nabla L({\mv w},\cal D) - \nabla L({\mv v}, \cal D)} \| \le {\rho}  \| {{\mv w} - {\mv v}} \|,
	\\
	\| {\nabla {L}({\hat{\mv w}}, \cal B) - \nabla {L}({\hat{\mv v}}, \cal B) } \| \le {\hat \rho } \| {{\hat{\mv w}} - {\hat{\mv v}}} \|.
\end{align}
Also, we have
\begin{align}
{L}({\mv w}) - {L}({\mv v}) \le \nabla {L}{({\mv v}, \cal D)^{\rm T}}({\mv w} - {\mv v}) + \frac{{{{\rho} }}}{2}\|{\mv w} - {\mv v}\|{^2},
\\
L({\hat{\mv w}}) - L({\hat{\mv v}}) \le \nabla L{({\hat{\mv v}}, \cal B)^{\rm T}}({\hat{\mv w}} - {\hat{\mv v}}) + \frac{{{\hat \rho }}}{2}\|{\hat{\mv w}} - {\hat{\mv v}}\|{^2}.
\end{align}

It follows from Assumption 2 that the gradients would not change arbitrarily quickly w.r.t. the model parameters in both pre-training and fine-tuning stages. 

{\bf Assumption 3} (Variance bound): The gradient estimate ${\nabla}{\tilde{L}}(\mv{w},\tilde{\mathcal D})$ (or ${\nabla}\tilde{L}_k(\hat {\mv{w}},\tilde{{\mathcal B}_k})$) of ${\nabla}{L}(\mv{w},\cal D)$ (or ${\nabla}L(\hat{\mv{w}},\cal B)$) on a mini-batch of data $\tilde{\mathcal D}$ (or $\tilde{\mathcal B}_k$) is assumed to be independent and unbiased for pre-training and fine-tuning stages, i.e., 
\begin{align}
	\mathbb{E}[{\nabla}{\tilde{L}}(\mv{w},\tilde{\mathcal D})]={\nabla}{L}(\mv{w},\cal D), \\
	\mathbb{E}[{\nabla}\tilde{L}_k(\hat {\mv{w}},\tilde{\mathcal B}_k)]={\nabla}{L}(\hat {\mv{w}}, \cal B).
\end{align}
Furthermore, the variance of ${\nabla}{\tilde{L}}(\mv{w},\tilde{\mathcal D})$ (or ${\nabla}\tilde{L}_k(\hat {\mv{w}},\tilde{\mathcal B}_k)$) is bounded by 
\begin{align}
\mathbb{E}\left[\|{\nabla}{\tilde{L}}(\mv{w},\tilde{\mathcal D})- {\nabla}{L}(\mv{w}, {\cal D})\|^2\right] \le {{\alpha}^2}/{|\tilde{\mathcal D}|},
\\
\mathbb{E}\left[\|{\nabla}\tilde{L}_k(\hat {\mv{w}},\tilde{\mathcal B}_k)- {\nabla} L(\hat{\mv{w}}, {\cal B})\|^2\right] \le {\hat \alpha^{2}}/{|\tilde{\mathcal B}_k|},
\end{align}
where $\alpha$ and $\hat \alpha$ are non-negative constants.

{\bf Assumption 4} (Lipschitz continuity w.r.t. the distance between data samples): For any two data samples ${d}_1 \in {\mathcal D}_1$ and ${d}_2 \in {\mathcal D}_2$ with diverse distributions ${\mathcal P}_1$ and ${\mathcal P}_2$, define $dist(\cdot,\cdot)$ as a distance function over  spaces ${\mathcal D}_1 \times {\mathcal D}_2$. We assume that for any $\mv w$, the loss ${\tilde{l}}(\mv{w},d)$ is Lipschitz w.r.t.  $d$ and the distance $dist(\cdot,\cdot)$, i.e.,
\begin{align}
	\| {l({\mv w},d_1) - l({\mv w},d_2)} \| \le {\tilde{\rho} } ~ dist(d_1,d_2).
\end{align}

Assumption 4 implies that the loss values considering the same model parameters ${\mv w}$ would not change arbitrarily large w.r.t. different data samples. Such an assumption could facilitate the convergence analysis considering the influence of potential variations in  data distributions across pre-training and fine-tuning stages.

{\bf Definition 1} (Wasserstein distance \cite{VPanaretos}): For two probability distributions ${\mathcal P}_1$ and ${\mathcal P}_2$, the Wasserstein distance between them is defined as 
\begin{align}
	W({\mathcal P}_1,{\mathcal P}_2)=\mathop {\inf }\limits_{\tilde{{\mathcal P}} \in \Pi ({\mathcal P}_1,{\mathcal P}_2)} {{\mathbb E}_{(d_1,d_2)\sim \tilde{{\mathcal P}}}}[dist(d_1,d_2)],
\end{align}
where $\Pi ({\mathcal P}_1,{\mathcal P}_2)$ denotes the set of all possible joint distributions with marginals ${\mathcal P}_1$ and ${\mathcal P}_2$ on the first and second coordinate. 
\begin{remark}
	It is worth noticing that we adopt the Wasserstein distance in this paper, as it has the following advantages over other distance measurements such as Kullback-Leibler (KL) divergence and Jensen-Shannon (JS) divergence. On the one hand, different from KL divergence and JS divergence, Wasserstein distance has the properties of positive definiteness, symmetry, and triangle inequality, and thus is a more appropriate measurement of the distance between two distributions. On the other hand, Wasserstein distance could measure the distance between any two distributions, while KL divergence and JS divergence require the distributions having the same support. Thus, Wasserstein distance is widely used to support theoretical analysis in existing works and has motivated many famous AI algorithms, such as Wasserstein Generative Adversarial Networks (WGAN) \cite{MArjovsky}.
\end{remark}
\vspace{-6pt}
\subsection{Convergence Analysis}
Under the above assumptions, we analyze the convergence behavior of the proposed two-stage edge learning system in terms of an upper bound on the average squared gradient norm. First, we consider  the second stage for  task-specific fine-tuning, during which the total expected improvement of loss function is provided in the following lemma.
\begin{lemma}\label{lemma1}
After $N$ rounds of task-specific fine-tuning on the initial model $\hat{\mv w}^{(0)}$, the total expected improvement of loss during model fine-tuning is
\begin{align} \label{summed improve finetune expectation} 
	&{\mathbb E}\left[{L}(\hat{\mv w}^{(N)},\cal B)\right] - {\mathbb E}\left[{L}(\hat{\mv w}^{(0)}, \cal B) \right]  \le \nonumber \\
	& \sum\limits_{n = 0}^{N-1}  \!\!\!- \hat{\gamma} (1 \!-\! \frac{{\hat{\rho} \hat{\gamma}}}{2}) {\mathbb E}\left[ \| \nabla { L}(\hat{\mv w}^{(n)}\!\!, \!{\cal B}) \|^2 \!\right] \!\!+\!\! \frac{{\hat{\rho} \hat{\alpha}^2 K \hat{\gamma}^2}}{2}  \!\! \sum\limits_{n = 0}^{N-1} \!\! \frac{1 } {{\tilde{B}^{(n)}_{\rm tot}}}.\!
\end{align}
\end{lemma}
\begin{IEEEproof}
	See Appendix A.
\end{IEEEproof}

Next, we focus on  the first stage for pre-training. We  analyze  ${\mathbb E}\left[{L}(\hat{\mv w}^{(0)}, \cal B) \right]$ in \eqref{summed improve finetune expectation}, or equivalently ${\mv w}^{(M)}$ as $\hat{\mv w}^{(0)}={\mv w}^{(M)}$. The following lemma provides an upper bound on ${\mathbb E}\left[{L}(\hat{\mv w}^{(0)},\cal B) \right]$.
\begin{lemma}\label{lemma2}
Suppose that the initial model is  $\mv{w}^{(0)}$ at the pre-training stage, and the distributions of datasets at pre-training and fine-tuning stages are given by $\cal P$ and $\hat{\cal P}$, respectively. It follows that after  $M$ rounds of model pre-training,  ${\mathbb E}\left[{L}(\hat{\mv w}^{(0)}, \cal B) \right]$ is upper bounded by
\vspace{-5pt}
\begin{align} \label{summed improve pretrain expectation}
	&{\mathbb E}\left[{L}(\hat{\mv w}^{(0)},\cal B) \right] \le - \sum\limits_{m = 0}^{M-1}  \gamma  (1-\frac{{\rho \gamma }}{2} ) {\mathbb E} \left[ \| \nabla L({\mv w}^{(m)},{\cal D}) \|^2 \right] \nonumber \\
	& \quad + \frac{{\rho \alpha^2 \gamma^2}}{2} \! \sum\limits_{m = 0}^{M-1} \! \!\frac{1} {\tilde{D}^{(m)}}\!+\!{\mathbb E}\left[L({\mv w}^{(0)},\cal D)\right]+\tilde{\rho}W(\hat{\mathcal{P}},\mathcal{P}).
\end{align}
\end{lemma}
\begin{IEEEproof}
	See Appendix B.
\end{IEEEproof}

Finally, by substituting Lemma \ref{lemma2} into Lemma \ref{lemma1} and with proper manipulation, we have the following theorem. 

\begin{theorem}\label{convergence}
Consider the two-stage edge learning with the model satisfying Assumptions 1$\sim$4, where the learning rate is given by $\bar \gamma= \gamma=\hat \gamma \le \min\{1/\rho, 1/\hat \rho\}$, and the initial model at the pre-training stage is $\mv{w}^{(0)}$. It follows that after $M$ rounds of model pre-training and $N$ rounds of task-specific  fine-tuning on datasets with distributions $\cal P$ and $\hat{\cal P}$, respectively, the average squared gradient norm has the following  upper bound:
\vspace{-2pt}
\begin{align} \label{convergence analysis}
	&\mathbb{E} \! \left[\! \frac{1}{(M+N)}\!\!\!\!\!\! \sum\limits_{r = 0}^{M+N-1}\!\!\!\!\! \| \nabla {L}({\mv w}^{(r)}, {\cal D}^{(r)}) \|^2 \! \right] \! \le \! \frac{\bar{\gamma}({\rho }{\alpha ^2}\Lambda + {{\hat \rho }{\hat \alpha ^2}}K \Omega)}{(M+N)} \nonumber \\
	&\qquad + \frac{2\tilde{\rho}W(\hat{\mathcal{P}},\mathcal{P})}{(M+N) \bar \gamma}+ \frac{2\left(L({\mv w}^{(0)},\cal D)-\hat{L}_{\rm inf}\right)}{(M+N) \bar \gamma} \nonumber \\
	& \qquad \buildrel \Delta \over = \Upsilon \big( M,N,\{\tilde{D}^{(m)}\},\{\tilde{B}^{(n)}_{k}\} \big),
	\end{align}
	\vspace{-3pt}where $\nabla {L}({\mv w}^{(r)}, {\cal D}^{(r)}) =  \nabla {L}({\mv w}^{(m)}, {\cal D})$ for $ 0 \le r \le M-1$, $\nabla {L}({\mv w}^{(r)}, {\cal D}^{(r)}) =  \nabla {L}({\hat {\mv w}}^{(n)}, {\cal B})$ for $M \le r \le M+N-1$, $\Lambda = \sum\nolimits_{m = 0}^{M - 1} {1}/ {\tilde{D}^{(m)}} $, and $\Omega= \sum\nolimits_{n = 0}^{N - 1} 1 /{\tilde{B}^{(n)}_{\rm tot}} $.
\end{theorem}
\begin{remark}
From Theorem \ref{convergence}, we have the following observations.
\begin{itemize}
	\item {\bf The upper bound $\Upsilon \big( M,N,\{\tilde{D}^{(m)}\},\{\tilde{B}^{(n)}_{k}\} \big)$ in Theorem \ref{convergence} shows that the pre-training and fine-tuning lead to different convergence behaviors depending on the number of learning rounds and batch sizes, which  provides useful  theoretical guidance on the system design.} In particular, the number of learning rounds (i.e., $M$ and $N$) and batch sizes (i.e., $\{\tilde{D}^{(m)}\}$ and $\{\tilde{B}^{(n)}_{k}\}$) of pre-training and fine-tuning stages jointly influence the convergence rate. Specifically, increasing  learning rounds $M$ and $N$ leads to more iterations of gradient descent to better fit the training data, while  increasing the batch sizes $\{\tilde{D}^{(m)}\}$ and $\{\tilde{B}^{(n)}_{k}\}$ results in more training data. Therefore, both of them could decrease the average gradient bound and increase the convergence rate. However, this in turn increases the overall learning delay and energy consumption. It is thus  crucial to properly design $M$, $N$, $\{\tilde{D}^{(m)}\}$, and $\{\tilde{B}^{(n)}_{k}\}$ jointly with other communication and computation resources for system performance enhancement under delay and energy consumption requirements. 
	\item {\bf The ultimate  performance of the target task  converges eventually as $N \rightarrow \infty$, with the average gradient bound landing on a finite limit.} When $M$, $\{\tilde{D}^{(m)}\}$, and $\{\tilde{B}^{(n)}_{k}\}$ are fixed, we have $\Upsilon \big( M,N,\{\tilde{D}^{(m)}\},\{\tilde{B}^{(n)}_{k}\} \big) \xrightarrow{N \rightarrow \infty} {\bar \gamma}{\hat \rho }{\hat \alpha ^2}K/{\tilde{B}_{\rm tot}}$, where ${\tilde{B}_{\rm tot}}$ is the fixed value of $\tilde{B}^{(n)}_{\rm tot} $'s. If $\hat \alpha =0$ or equivalently the full gradient descent is considered with zero noise on the gradient estimation, then the average gradient bound diminishes to zero, i.e., $\| \nabla {L}({\hat{\mv w}}^{(n)},{\cal B}) \|\xrightarrow{N \rightarrow \infty} 0$. By contrast, if $\hat \alpha >0$, i.e., the gradient estimation noise exists, then the limit of the average gradient bound ${\bar \gamma}{\hat \rho }{\hat \alpha ^2}K/{\tilde{B}_{\rm tot}}$ is non-zero. In this case, the finite bound ${\bar \gamma}{\hat \rho }{\hat \alpha ^2}K/{\tilde{B}_{\rm tot}}$ indicates that the gradients become relatively small as the training process keeps on, which implies the final converged behavior of the SGD method. 	 
\end{itemize}
\end{remark}

\section{Joint Communication and Computation Resource Management for Two-Stage Edge Learning}

\subsection{Problem Formulation}
This section jointly designs the communication and computation resource management over both pre-training and fine-tuning stages to optimize the edge learning performance while ensuring the energy and delay requirements. Specifically, our objective is to minimize the average squared gradient norm bound in \eqref{convergence analysis}, by jointly optimizing the number of learning rounds of pre-training and fine-tuning $M$ and $N$, batch sizes $\{\tilde{D}^{(m)}\}$ and $\{\tilde{B}^{(n)}_{k}\}$, clock frequencies $\{f^{(m)}\}$ and $\{{\hat f}^{(n)}_k\}$, as well as the power control schemes of edge devices for gradient uploading $\{p^{(n)}_{k}\}$, subject to a series of transmit power constraints. The formulation also takes into account the requirements on the total energy consumption budget  and the overall system learning delay. By denoting the resource allocation variables as $\Psi \buildrel \Delta \over = \big\{\{\tilde{D}^{(m)}\}, \{\tilde{B}^{(n)}_{k}\},$ $ \{{ f}^{(m)}\}, \{{\hat f}^{(n)}_k\}, \{p^{(n)}_{k}\}\big\}$, the learning performance optimization problem is formulated as
\begin{subequations}\label{P1}
	\begin{align}
	\text{(P1)}: \mathop {\min }\limits_{M,N,\Psi} &  \Upsilon \big( M,N,\{\tilde{D}^{(m)}\},\{\tilde{B}^{(n)}_{k}\} \big) \nonumber\\
	\mathrm{s.t.}~&
	0 \le \tilde{D}^{(m)} \le D^{\rm max}, \forall m \label{P1-a}\\
	~
	&0 \le \tilde{B}^{(n)}_{k} \le B^{\rm max}_k, \forall n,k \label{P1-b}\\
	~
	&0 \le f^{(m)} \le f^{\rm max}, \forall m \label{P1-c}\\
	~
	&0 \le {\hat f}^{(n)}_k \le {\hat f}_k^{\rm max}, \forall n,k \label{P1-d}\\
	~
	&\tilde{\tau}\left(M,N,\{\tilde{D}^{(m)}\},\{f^{(m)}\},\{\tilde{B}^{(n)}_{k}\},\right. \nonumber \\
	& \qquad \left. \{{\hat f}^{(n)}_k\},\{p^{(n)}_{k}\}\right) \le \tilde \tau_0 \label{P1-f}\\
	~
	&{\tilde E}\left( M,N,\{\tilde{D}^{(m)}\},\{f^{(m)}\},\{\tilde{B}^{(n)}_{k}\},\right. \nonumber \\
	& \qquad \left. \{{\hat f}^{(n)}_k\},\{p^{(n)}_{k}\} \right) \le \tilde E_0 \label{P1-g}\\
	~
	&(10), (11),\nonumber
	\end{align}
	\end{subequations}
where $D^{\rm max}$, $B^{\rm max}_k$, $f^{\rm max}$, and ${\hat f}_k^{\rm max}$ in \eqref{P1-a}$\sim$\eqref{P1-d} denote the maximum batch sizes and clock frequencies for pre-training and fine-tuning stages, $\tilde \tau_0$ and $\tilde E_0$ denote the thresholds of overall learning delay and energy consumption requirements, respectively. By properly adjusting $\tilde \tau_0$ and $\tilde E_0$, we can  balance the trade-off among the convergence rate for two-stage training as well as the energy consumption and training delay. However, in problem (P1), the objective is  non-convex, as learning rounds $M$ and $N$ are closely coupled with batch sizes $\{\tilde{D}^{(m)}\}$ and $\{\tilde{B}^{(n)}_{k}\}$. The constraints in \eqref{P1-f} and \eqref{P1-g} are also highly non-convex, due to the close coupling  of  variables $M$, $N$, and $\Psi$. Furthermore, the integer variables $M$ and $N$ make problem (P1) a mixed-integer non-linear problem (MINLP). Therefore,  problem (P1) is highly non-convex and non-trivial to be optimally solved in general.

\subsection{Proposed Solution to Problem (P1)}
We propose an effective algorithm to solve problem (P1). In particular, we first use the SCA technique to optimize $\Psi$ in (P1) under given number of learning rounds $M$ and $N$, and then find the optimized $M$ and 
$N$ achieving the minimum objective value via a two-dimensional search. In the following, we focus on optimizing $\Psi$ under given $M$ and $N$. 

With fixed $M$ and $N$, the objective of problem (P1) becomes convex w.r.t batch sizes $\{\tilde{D}^{(m)}\}$ and $\{\tilde{B}^{(n)}_{k}\}$. To start with, we introduce auxiliary variables $\{\tilde{D}^{(m)\prime}\}$ and $\{\tilde{B}^{(n)\prime}_{k}\}$, then the optimization of $\Psi$ in problem (P1) with given $M$ and $N$ is reformulated as 
\begin{subequations}\label{P2}
	\begin{align}
	\text{(P2)}:& \mathop {\min } \limits_{\{\tilde{D}^{(m)\prime}\},\{\tilde{B}^{(n)\prime}_{k}\}, \Psi}  \Upsilon \big( M,N,\{\tilde{D}^{(m)}\},\{\tilde{B}^{(n)}_{k}\} \big) \nonumber\\
	& \mathrm{s.t.}~
	 \sum\limits_{m = 0}^{M-1}  \frac{N_{\rm FLOP} }{cf^{(m)} \tilde{D}^{(m)\prime} } + \sum\limits_{n = 0}^{N-1} \mathop {\max }\limits_{k \in {\cal K}} \Bigg(\frac{\beta}{{r}^{(n),\rm d}_{k}}+ \nonumber \\
	&  \qquad   \frac{ N_{\rm FLOP}} {\hat{c}_k \hat{f}_k^{(n)}  \tilde{B}^{(n)\prime}_{k}}\!+\! \frac{{\beta}}{W_k^{\rm u} \log_2 \! \left(1 \!+\!\frac{g^{(n),\rm u}_{k} p^{(n)}_{k}}{ W_k^{\rm u}N_0^{(n)}} \right)}\Bigg) \!\le\! \tilde \tau_0 \label{P2-a}\\
	~
	&\qquad \sum\limits_{m = 0}^{M-1}  \eta \frac{ N_{\rm FLOP}\phi {f^{(m)}}^2  }{c\tilde{D}^{(m)\prime}}+ \nonumber \\
	&\qquad  \sum\limits_{n = 0}^{N-1} \!\! \left( \!\tilde P \!\times\! \mathop {\max }\limits_{k \in {\cal K}} (\frac {\beta}{{r}^{(n),\rm d}_{k}}) \!+\! \sum\limits_{k = 1}^K \left(   \frac{\hat{\eta}_k {N_{\rm FLOP} } \hat{\phi}_k {{\hat f}_k^{(n)2}}} {\hat{c}_k \tilde{B}^{(n)\prime}_{k}} \right. \right. \nonumber \\
	&\qquad \left. \left. + \frac{p^{(n)}_{k}{\beta}}{W_k^{\rm u} \log_2 \left(1+\frac{g^{(n),\rm u}_{k} p^{(n)}_{k}}{ W_k^{\rm u}N_0^{(n)}} \right)} \right)\right)  \le \tilde E_0 \label{P2-b}\\
	~
	& \qquad \tilde{D}^{(m)} \le \frac{1}{\tilde{D}^{(m)\prime}}, \forall m \label{P2-c}\\
	~
	& \qquad \tilde{B}^{(n)}_{k} \le \frac{1}{\tilde{B}^{(n)\prime}_{k}}, \forall k,n \label{P2-d}\\
	~
	& \qquad (10), (11), (35 \rm a) \sim (35 \rm d).\nonumber
	\end{align}
	\end{subequations}
It is worth noticing that the objective in problem (P2) is convex and constraint \eqref{P2-a} is also convex w.r.t. $\big\{\{\tilde{D}^{(m)\prime}\}, \{\tilde{B}^{(n)\prime}_{k}\}, \{{ f}^{(m)}\}, \{{\hat f}^{(n)}_k\}, \{p^{(n)}_{k}\}\big\}$. However, problem (P2) is still  non-convex  due to the non-convex constraints \eqref{P2-b} $\sim$ \eqref{P2-d}.

Next, we use the SCA technique to address non-convex constraints \eqref{P2-b} $\sim$ \eqref{P2-d} by implementing the first-order Taylor expansion in an iterative manner. Consider each iteration $o \ge 1$, where the local points are denoted by $\big\{\{\tilde{D}^{(m)\prime^{(o)}}\}, \{\tilde{B}^{(n)\prime^{(o)}}_{k}\}, \{p^{(n)^{(o)}}_{k}\}\big\}$. Then the last term of the left hand side of \eqref{P2-b} is upper bounded by 
\begin{align} \label{p approximation}
&\frac{p^{(n)}_{k}{\beta}}{W_k^{\rm u} \log_2 \left(1+\frac{g^{(n),\rm u}_{k} p^{(n)}_{k}}{ {W_k^{\rm u}N_0^{(n)}}} \right)}  \nonumber \\
&\le \frac{p^{(n)^{(o)}}_{k}{\beta}}{W_k^{\rm u} \log_2 \left(1+\frac{g^{(n),\rm u}_{k} p^{(n)^{(o)}}_{k}}{ W_k^{\rm u}N_0^{(n)}} \right)} +  u(p^{(n)^{(o)}}_{k})(p^{(n)}_{k}-p^{(n)^{(o)}}_{k}) \nonumber \\
&\buildrel \Delta \over = \zeta^{(o)}(p^{(n)}_{k}),
\end{align}
where 
\begin{align} \label{u in p}
&u(p^{(n)^{(o)}}_{k})=\frac{\beta}{W_k^{\rm u} \log_2 \left(1+\frac{g^{(n),\rm u}_{k} p^{(n)^{(o)}}_{k}}{ W_k^{\rm u}N_0^{(n)}} \right)}- \nonumber \\
&\frac{p^{(n)^{(o)}}_{k}{\beta}g^{(n),\rm u}_{k}}{W_k^{\rm u} \log_2\! \left( \!1\!+\!\frac{g^{(n),\rm u}_{k} p^{(n)^{(o)}}_{k}}{ W_k^{\rm u}N_0^{(n)}}\! \right)^2 \!\!\!( W_k^{\rm u}N_0^{(n)}\!+\!g^{(n),\rm u}_{k} p^{(n)^{(o)}}_{k}\!)\!\ln2}.
\end{align}
Furthermore, we deal with the non-convex constraints in \eqref{P2-c} and \eqref{P2-d} by approximating them based on their first-order Taylor expansions, i.e., 
\begin{align} 
&\tilde{D}^{(m)} - \frac{1}{\tilde{D}^{(m)\prime}} \le \nonumber \\
& \tilde{D}^{(m)} \!-\! \frac{1}{\tilde{D}^{(m)\prime^{(o)}}}\!+\!\frac{1}{\tilde{D}^{(m)\prime^{(o)2}}}(\tilde{D}^{(m)\prime}\!\!-\!\!\tilde{D}^{(m)\prime^{(o)}}) \!\le\! 0, \!\forall m, \label{29c approximation} \\
&\tilde{B}^{(n)}_{k}- \frac{1}{\tilde{B}^{(n)\prime}_{k}} \le \nonumber \\
&\tilde{B}^{(n)}_{k}\!-\!\frac{1}{\tilde{B}^{(n)\prime^{(o)}}_{k}}\!+\!\frac{1}{\tilde{B}^{(n)\prime^{(o)2}}_{k}}(\tilde{B}^{(n)\prime}_{k}\!-\!\tilde{B}^{(n)\prime^{(o)}}_{k}) \!\le\! 0, \!\forall k,n. \label{29d approximation}
\end{align}

Then, by replacing the last term of the left hand side of \eqref{P2-b} as $\zeta^{(o)}(p^{(n)}_{k})$ in \eqref{p approximation}, and replacing \eqref{P2-c} and \eqref{P2-d} as \eqref{29c approximation} and \eqref{29d approximation}, respectively. We obtain the approximate convex version
of problem (P2) in the $o$-th iteration as problem (P3.$o$) under fixed $M$ and $N$, which can be efficiently solved via standard convex optimization tools such as CVX \cite{MG2014}.
\begin{subequations}\label{P3}
	\begin{align}
	{({\rm P}3.o)}:& \mathop {\min } \limits_{\{\tilde{D}^{(m)\prime}\},\{\tilde{B}^{(n)\prime}_{k}\}, \Psi}  \Upsilon \big( M,N,\{\tilde{D}^{(m)}\},\{\tilde{B}^{(n)}_{k}\} \big) \nonumber\\
	& \mathrm{s.t.}
	\sum\limits_{m = 0}^{M-1}  \eta \frac{ N_{\rm FLOP}\phi {f^{(m)}}^2  }{c\tilde{D}^{(m)\prime}}\!\!+\!\! \sum\limits_{n = 0}^{N-1} \left( \tilde P \times \mathop {\max }\limits_{k \in {\cal K}} (\frac {\beta}{{r}^{(n),\rm d}_{k}}) \right. \nonumber\\
	& \left. + \sum\limits_{k = 1}^K \left(   \frac{\hat{\eta}_k {N_{\rm FLOP} } \hat{\phi}_k {{\hat f}_k^{(n)2}}} {\hat{c}_k \tilde{B}^{(n)\prime}_{k}} + \zeta^{(o)}(p^{(n)}_{k}) \right)\right) \le \tilde E_0 \label{P3-b}\\
	~
	& \qquad \tilde{D}^{(m)}-\frac{1}{\tilde{D}^{(m)\prime^{(o)}}}+\nonumber \\
	&\qquad \quad \frac{1}{\tilde{D}^{(m)\prime^{(o)2}}}(\tilde{D}^{(m)\prime}-\tilde{D}^{(m)\prime^{(o)}}) \le 0, \forall m \label{P3-c}\\
	~
	& \qquad \tilde{B}^{(n)}_{k}-\frac{1}{\tilde{B}^{(n)\prime^{(o)}}_{k}}\nonumber \\
	& \qquad \quad +\frac{1}{\tilde{B}^{(n)\prime^{(o)2}}_{k}}(\tilde{B}^{(n)\prime}_{k}-\tilde{B}^{(n)\prime^{(o)}}_{k}) \le 0, \forall k,n \label{P3-d}\\
	~
	& \qquad (10),(11), (35 \rm a) \sim (35 \rm d), (36 \rm a).\nonumber
	\end{align}
	\end{subequations}
For each fixed $M$ and $N$, we iteratively solve problem (P3.$o$) to obtain a series of solutions $\big\{\{\tilde{D}^{(m)\prime^{(o)}}\}, \{\tilde{B}^{(n)\prime^{(o)}}_{k}\}, \Psi^{(o)}\big\}$'s, which lead to monotonically non-increasing objective values for (P2). As a result, the convergence of the SCA-based algorithm to solve problem (P2)
is ensured. Such iterations stop until the decrease of the objective value is smaller than a given threshold $\epsilon$. We summarize the proposed algorithm for solving problem (P1) in Algorithm 1.

\begin{algorithm}[h]
	\caption{Overall Algorithm for solving problem (P1)}
	\label{alg1}
	\begin{algorithmic}[1] 
	\State Initialize an empty buffer $\Xi$ to store the optimal value.
	\For{$M=1:M^{\rm max}$}
	\For{$N=1:N^{\rm max}$}
	\State Initialize the auxiliary variables $\{\tilde{D}^{(m)\prime^{(0)}}\}$ and $\{\tilde{B}^{(n)\prime^{(0)}}_{k}\}$, and the transmit power $\{p^{(n)^{(0)}}_{k}\}$; Set $o=1$.
	\Repeat
	\State Solve problem (P3.$o$) under local points $\big\{ \{\tilde{D}^{(m)\prime^{(o-1)}}\}, \{\tilde{B}^{(n)\prime^{(o-1)}}_{k}\},\{p^{(n)^{(o-1)}}_{k}\} \big\}$ to obtain solutions $\big\{ \{\tilde{D}^{(m)\prime^{(o)*}}\}, \{\tilde{B}^{(n)\prime^{(o)*}}_{k}\},\Psi^{(o)*} \big\}$. 
	\State Update $\big\{ \{\tilde{D}^{(m)\prime^{(o)}}\}, \{\tilde{B}^{(n)\prime^{(o)}}_{k}\},\{p^{(n)^{(o)}}_{k}\} \big\}=\big\{ \{\tilde{D}^{(m)\prime^{(o)*}}\}, \{\tilde{B}^{(n)\prime^{(o)*}}_{k}\},\{p^{(n)^{(o)*}}_{k}\} \big\}$, $o=o+1$.
	\Until the decrease of the objective value is below a given threshold $\epsilon$. 
	\State Obtain final solutions $\big\{ \{\tilde{D}^{(m)\prime*}\}, \{\tilde{B}^{(n)\prime*}_{k}\}, \Psi^{*} \big\}$ of current $M$ and $N$. Round $\big\{ \{\tilde{D}^{(m)^{*}}\}, \{\tilde{B}^{(n)^{*}}_{k}\}\big \}$ to their nearby integers.
	\State Compute  $\Upsilon \big( M,N,\{\tilde{D}^{(m)^{*}}\},\{\tilde{B}^{(n)^{*}}_{k}\} \big)$.
	\If{$\Upsilon \big( M,N,\{\tilde{D}^{(m)^{*}}\},\{\tilde{B}^{(n)^{*}}_{k}\} \big)$ is smaller than the stored one in $\Xi$ {\emph {or}} $\Xi$ is empty}
	\State Replace the optimal solution as $\big\{M, N, \{\tilde{D}^{(m)\prime*}\}, \{\tilde{B}^{(n)\prime*}_{k}\}, \Psi^{*} \big\}$.
	\EndIf
	\EndFor
	\EndFor
	\State {\bf Output} the final optimal solution $\big\{M^*, N^*, \{\tilde{D}^{(m)\prime*}\},$\\ $\{\tilde{B}^{(n)\prime*}_{k}\}, \Psi^{*} \big\}$ in $\Xi$.
	\end{algorithmic}
\end{algorithm}

\section{Numerical Results}
This section presents numerical results to validate the performance
of the proposed two-stage edge learning system design with joint pre-training and fine-tuning. In the experiments, we adopt MNIST \cite{MNIST},  CIFAR-10 \cite{CIFAR10}, and CINIC-10 \cite{CINIC10} datasets to train LeNet \cite{YLecun} and GoogLeNet \cite{CSzegedy} for performance validation,  respectively. Specifically, in the first experiment, we pre-train and fine-tune LeNet on MNIST, while in the second experiment, we pre-train GoogLeNet on CINIC-10 and fine-tune it on CIFAR-10. We conduct experiments based on PyTorch with NVIDIA RTX 3090 GPUs. In the following, we introduce the datasets.
\begin{itemize}
	\item {\bf MNIST:} MNIST \cite{MNIST} contains images of handwritten digits each with 28 grey pixels. The number of training samples is 60,000, and the number of testing samples is 10,000.
	\item {\bf CIFAR-10:} CIFAR-10 \cite{CIFAR10} contains colorful
	images each with $32 \times 32 \times 3$ pixels of real-world
	objects from 10 classes (such as automobile and dog). It has a training set containing 50,000 samples, and a test set containing 10,000 samples.
	\item {\bf CINIC-10:} CINIC-10 \cite{CINIC10} collects 270,000 colorful images with size $32 \times 32 \times 3$, which are equally split into training, validating, and test sets. It has 10 same classes of objects as CIFAR-10, which are mostly sampled from ImageNet images.
\end{itemize}

\begin{figure*}[h]
	\vspace{-12pt}
	\centering
	\subfigure[MNIST to MNIST]{
	\centering
	\includegraphics[width=0.49\linewidth]{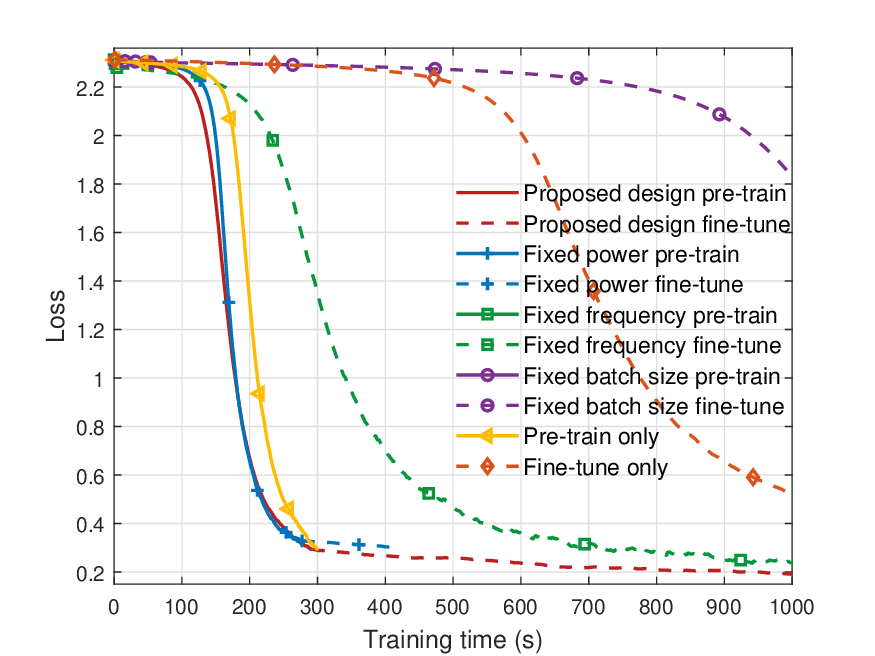}\label{loss_delay_a}
	}%
	\subfigure[CINIC-10 to CIFAR-10] {
	\centering
	\includegraphics[width=0.49\linewidth]{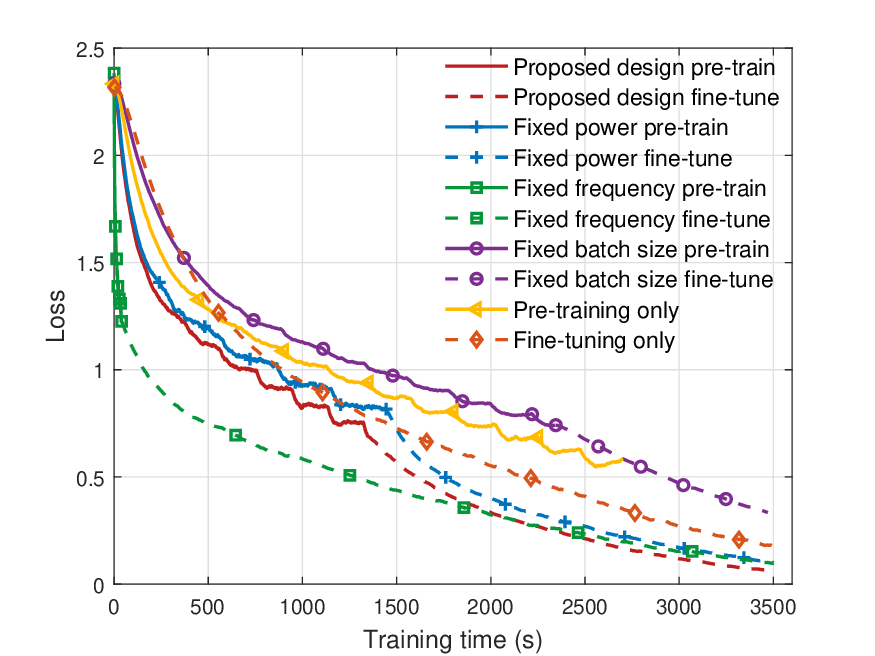}\label{loss_delay_b}
	}\vspace{-5pt}
	\caption{Convergence behavior in terms of  the training loss over the overall training time. (a) Pre-training and fine-tuning on MNIST. (b) Pre-training on CINIC-10 and fine-tuning on CIFAR-10.}
	\label{loss_delay}
	\vspace{-10pt}
\end{figure*}

\begin{figure*}[h]
	\vspace{-5pt}
	\centering
	\subfigure[MNIST to MNIST]{
	\centering
	\includegraphics[width=0.49\linewidth]{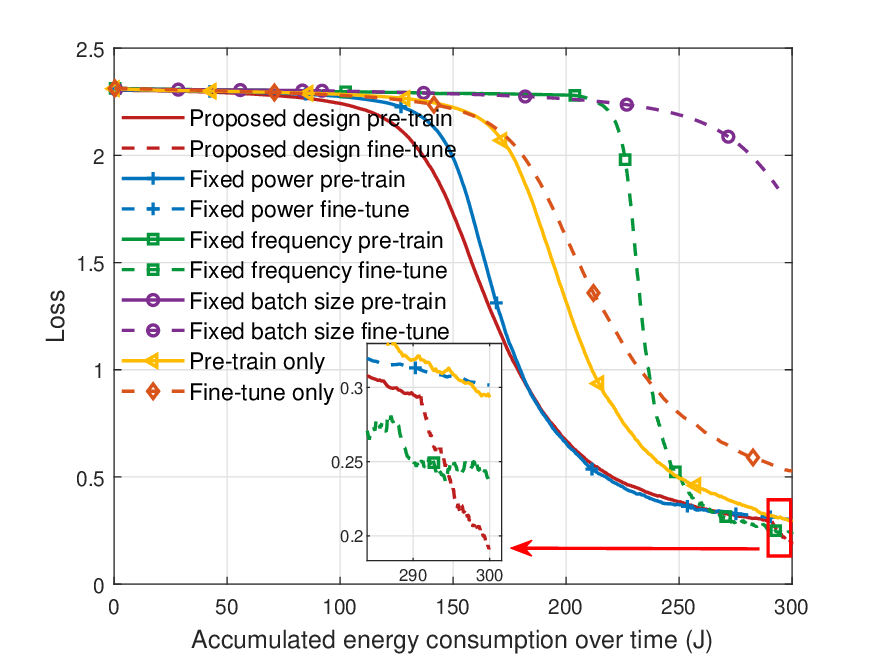}\label{loss_energy_a}
	}%
	\subfigure[CINIC-10 to CIFAR-10]{
	\centering
	\includegraphics[width=0.49\linewidth]{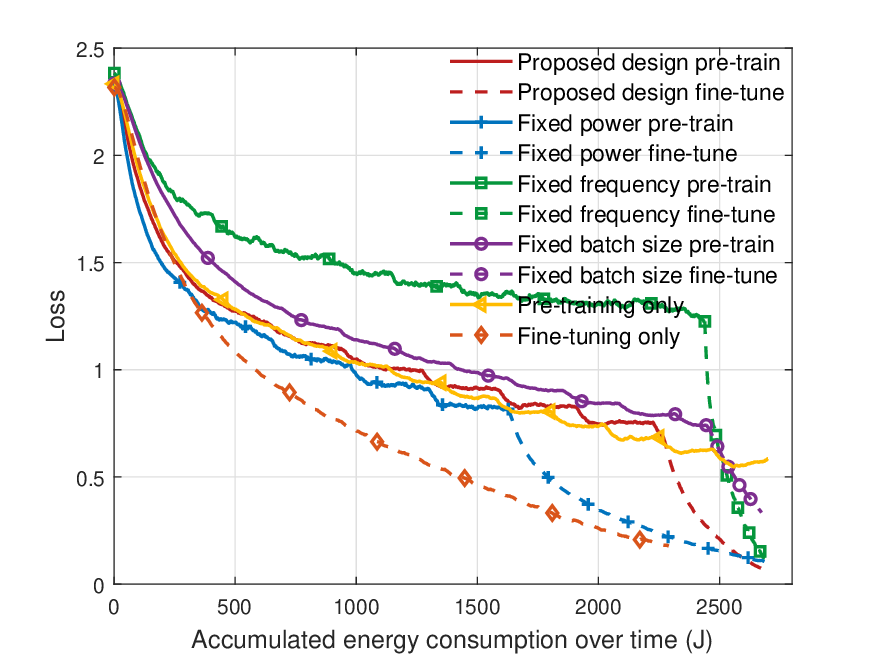}\label{loss_energy_b}
	}\vspace{-5pt}
	\caption{Convergence behavior in terms of the training loss w.r.t. the accumulated overall system energy consumption over time. (a) Pre-training and fine-tuning on MNIST. (b) Pre-training on CINIC-10 and fine-tuning on CIFAR-10.}
	\label{loss_energy}
	\vspace{-10pt}
\end{figure*}

In the simulation, we adopt similar parameter setups as in \cite{XDeng2023,MWu2021}. Specifically, we set the learning rate as $\bar \gamma=0.01$, the number of devices as $K = 3$, the transmit power of the edge server as $\tilde P= 0.5~{\rm W}$, the average power budgets and maximum power budgets of different devices as $P_k^{\rm ave}=P_k^{\rm max}=0.5~{\rm W},\forall k$. For wireless channels, we generate channel coefficients based on independently and identically (i.i.d.) Rayleigh fading with average path
loss of $10^{-5}$ and fix them across different communication rounds. For the \emph{MNIST to MNIST task}, we set the parameters as follows. Based on real measurement via PyTorch, LeNet has $61.71~{\rm K}$ parameters occupying $244.32~{\rm KB}$ of memory space. Its forward propagation workload is $416.52~{\rm KFLOPs}$. It is non-trivial to measure the workload of back propagation in practice. According to \cite{openai2018}, the total workload considering both forward and backward propagation is $2 \sim 3$ times of the forward one. As such, we set the model/gradient size as $\beta=0.2~{\rm Mbit}$ and the computation workload as $N_{\rm FLOP}=1~{\rm MFLOPs}$. For other communication parameters, we set the bandwidth as $W_k^{\rm u}=W_d=100~{\rm kHz},\forall k$, and the PSD of AWGN as 
$N_0^{(n)}=\tilde{N}_{0,k}^{(n)}=10^{-16}~{\rm W/Hz}, \forall k,n$. For computation parameters, we set the maximum clock frequencies as $f^{\rm max}=1600~{\rm MHz}$, ${\hat f}_k^{\rm max}=360~{\rm MHz}, \forall k$, the maximum sizes of data batch as $D^{\rm max}={B}_k^{\rm max}=700, \forall k$, the PUEs as $\eta=4$, ${\hat \eta}_k=1, \forall k$, the power coefficients as $\phi=3.91\times 10^{-27}~{\rm W/(cycle/s)^3}$, and $\{\phi_k\}=\{1.09, 1.56, 2.34\}\times 10^{-27}~{\rm W/(cycle/s)^3}$ \cite{XDeng2023}. Finally, we set the training delay threshold as $\tilde \tau_0=1000~{\rm s}$, and the energy consumption threshold as $\tilde E_0=300~{\rm J}$. For the \emph{CINIC-10 to CIFAR-10 task}, based on real measurement, GoogLeNet has $6.17~{\rm M}$ parameters occupying $24.29~{\rm MB}$ of memory space, and its forward computation workload is $1.53~{\rm GFLOPs}$. As such, we set $\beta=24~{\rm Mbit}$ and $N_{\rm FLOP}=4~{\rm GFLOPs}$. Moreover, we set $W_k^{\rm u}=W_d=3~{\rm MHz},\forall k$, and $N_0^{(n)}=\tilde{N}_{0,k}^{(n)}=3.3 \times 10^{-18}~{\rm W/Hz}, \forall k,n$. For computation parameters, we set $f^{\rm max}=4000~{\rm MHz}$, ${\hat f}_k^{\rm max}=900~{\rm MHz}, \forall k$, $D^{\rm max}={B}_k^{\rm max}=500, \forall k$, $\phi=2.5\times 10^{-28}~{\rm W/(cycle/s)^3}$, and $\{{\hat \phi}_k\}=\{0.7, 1,1.5\}\times 10^{-28}~{\rm W/(cycle/s)^3}$ \cite{MWu2021}. Finally, we set $\tilde \tau_0=3500~{\rm s}$ and $\tilde E_0=2700~{\rm J}$.

\begin{figure*}[h]
	\vspace{-10pt}
	\centering
	\subfigure[MNIST to MNIST]{
	\centering
	\includegraphics[width=0.49\linewidth]{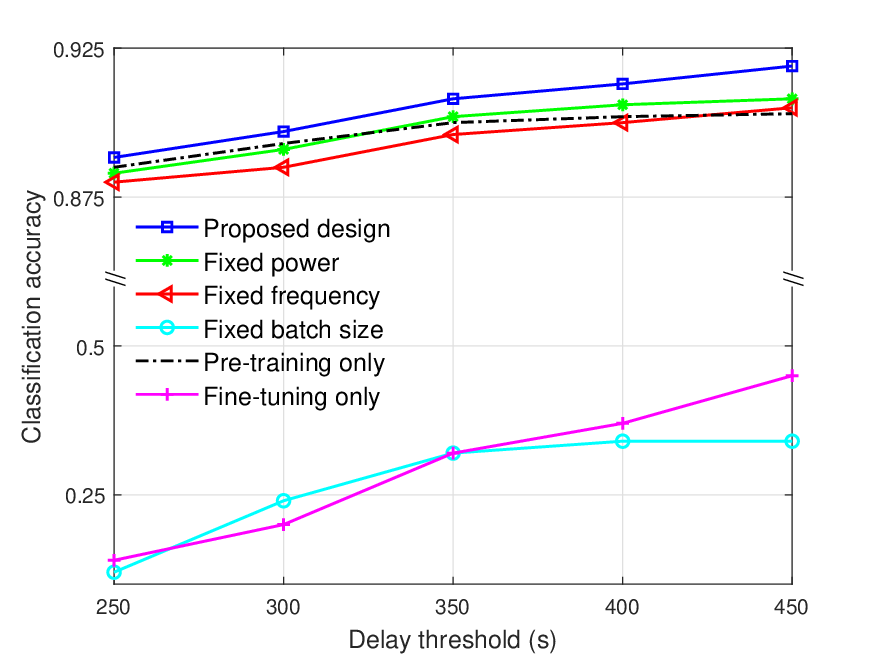}\label{acc_delay_a}
	}%
	\subfigure[CINIC-10 to CIFAR-10]{
	\centering
	\includegraphics[width=0.49\linewidth]{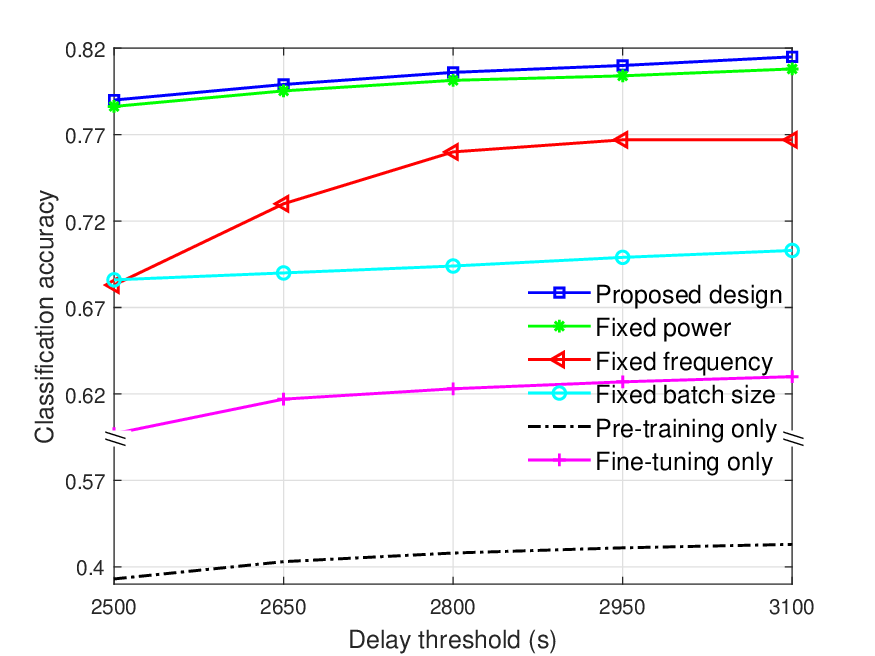}\label{acc_delay_b}
	}\vspace{-5pt}
	\caption{Classification accuracy w.r.t. different system delay thresholds. (a) Pre-training and fine-tuning on MNIST under the energy consumption threshold  $\tilde E_0 =250~ {\rm J}$. (b) Pre-training on CINIC-10 and fine-tuning on CIFAR-10 under the energy consumption threshold $\tilde E_0 =1700~ {\rm J}$.}
	\label{acc_delay}
	\vspace{-10pt}
\end{figure*}

\begin{figure*}[h]
	\centering
	\subfigure[MNIST to MNIST]{
	\centering
	\includegraphics[width=0.49\linewidth]{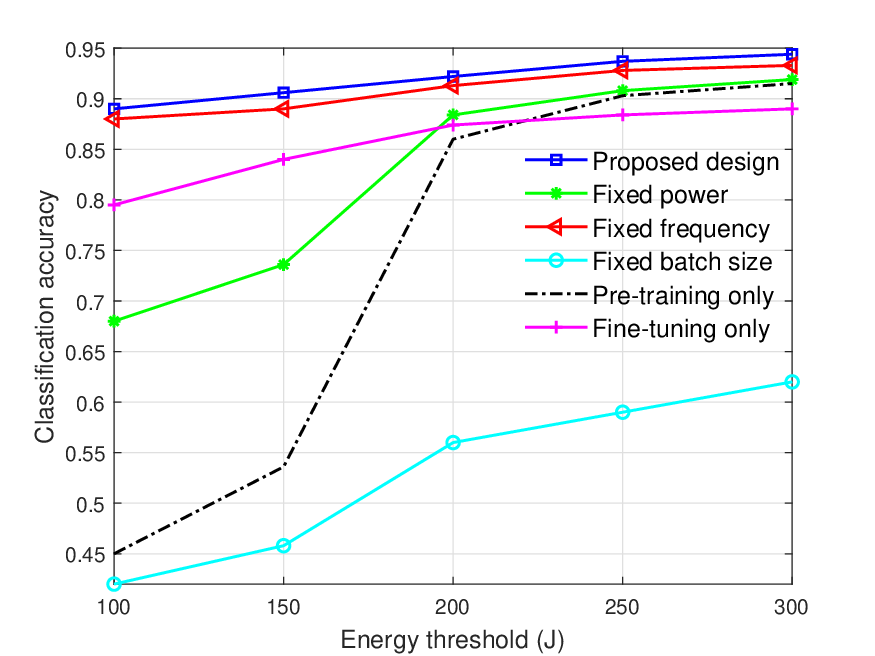}\label{acc_energy_a}
	}%
	\subfigure[CINIC-10 to CIFAR-10]{
	\centering
	\includegraphics[width=0.49\linewidth]{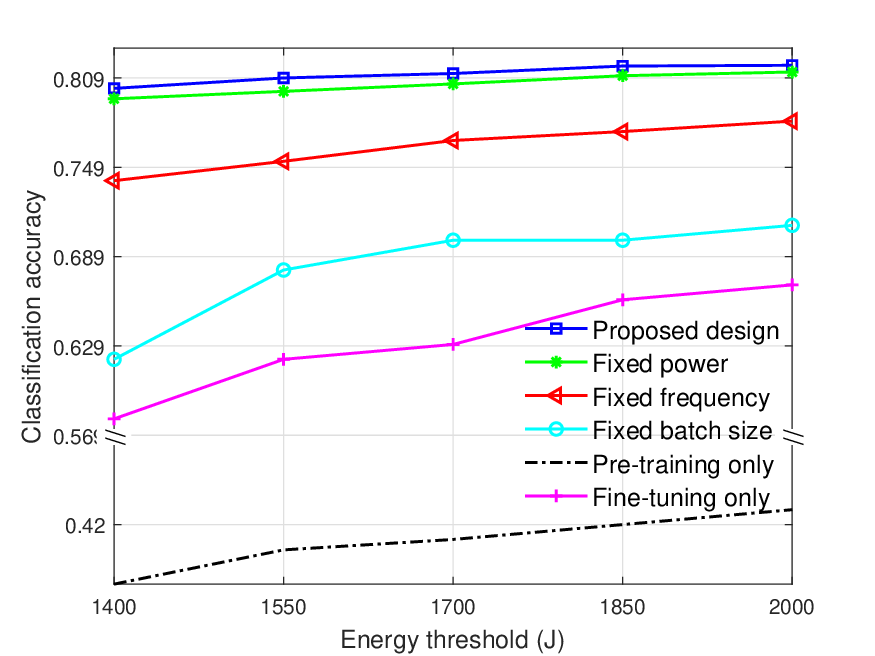}\label{acc_energy_b}
	} \vspace{-10pt}
	\caption{Classification accuracy w.r.t. different system energy consumption thresholds. (a) Pre-training and fine-tuning on MNIST under the training delay threshold $\tilde \tau_0 =1000~ {\rm s}$. (b) Pre-training on CINIC-10 and fine-tuning on CIFAR-10 under the training delay threshold  $\tilde \tau_0 =3000~ {\rm s}$.}
	\label{acc_energy}
	\vspace{-10pt}
\end{figure*}

The following benchmark schemes are considered for performance comparison. 
\begin{itemize}
	\item {\bf Fixed power design:} We fix the transmit power as the maximum transmit power, i.e., $p^{(n)}_{k} = {P}_k^{\rm max}, \forall n,k$, and then optimize the remaining variables in problem (P1) based on the proposed solution.
	\item {\bf Fixed clock frequency design:} We fix the clock frequencies as $f^{(m)} = f^{\rm max}, \forall m$, and ${\hat f}^{(n)}_k = {\hat f}_k^{\rm max}, \forall n,k$, and then optimize problem (P1).
	\item {\bf Fixed batch size design:} We fix the batch sizes as $\tilde{D}^{(m)} = D^{\rm max}, \forall m $, and 
	$\tilde{B}^{(n)}_{k} =  B^{\rm max}_k, \forall n,k$, and then optimize problem (P1). 
	\item {\bf Pre-training only:} We only pre-train the model without task-specific fine-tuning by leveraging the pre-stored data at the edge server in a centralized learning manner, i.e., solving problem (P1) by setting $N=0$.
	\item {\bf Fine-tuning only:} We only fine-tune the model from the scratch via FEEL without performing model pre-training. In this case, we perform task-specific fine-tuning with data at devices via FEEL, i.e., solving problem (P1) by setting $M=0$.
\end{itemize}

Figs. \ref{loss_delay_a} and \ref{loss_delay_b}  show the  convergence behavior in terms of the  training loss over the overall training time for the MNIST to MNIST and CINIC-10 to CIFAR-10 tasks, respectively. It is observed that, given a training delay threshold requirement, our proposed design outperforms other benchmark schemes and achieves the smallest loss value when the training delay threshold is met. It is worth noting that although the loss values in some benchmarks may decrease swiftly during the early training stages, the final converged performance of our proposed design surpasses the benchmarks, which shows that the proposed design better leverages communication and computation resources over  pre-training and fine-tuning stages to achieve better learning performance.

Figs. \ref{loss_energy_a} and \ref{loss_energy_b} show the  convergence behavior in terms of the training loss w.r.t. the accumulated overall system energy consumption over time for the MNIST to MNIST and CINIC-10 to CIFAR-10 tasks, respectively. It is observed that, under the given energy consumption threshold, our proposed design achieves better learning performance (lower loss values) than other benchmark schemes when meeting the energy consumption threshold. Although some benchmark schemes enable faster loss value decreasing in the early training stage, our proposed design shows superiority at the end of fine-tuning. Such observation verifies the effectiveness  of our joint pre-training and fine-tuning design.

Figs. \ref{acc_delay_a} and \ref{acc_delay_b} show the testing performance in terms of classification accuracy w.r.t. different system delay thresholds for the MNIST to MNIST and CINIC-10 to CIFAR-10 tasks, respectively. It is observed that our proposed design outperforms other benchmark schemes, which validates the benefits of  joint communication and computation design  in improving the overall learning efficiency. It is also  observed that, as training delay threshold becomes larger, the performance tends to converge. This is because, in such a case, system energy consumption acts as the main system bottleneck. Even though we enlarge the training delay threshold, the system performance could not be further improved with limited energy consumption budget. Furthermore, it is also observed that the performance of the pre-training only and fine-tuning only designs compromise compared with our proposed design, which shows the superiority of considering the entire life-cycle of edge learning with model pre-training and task-specific fine-tuning.

Figs. \ref{acc_energy_a} and \ref{acc_energy_b} show the classification accuracy w.r.t. system energy consumption thresholds for the MNIST to MNIST and CINIC-10 to CIFAR-10 tasks, respectively. It is observed that our proposed design outperforms other benchmark schemes for efficient learning. Furthermore, it is shown that, as energy consumption threshold getting larger, the learning performance also tends to converge. This is due to the fact that the limited system delay requirement becomes  the main system bottleneck in this case.

\begin{figure}[h]
	\vspace{-11pt}
	\centering
	\includegraphics[width=0.97\linewidth]{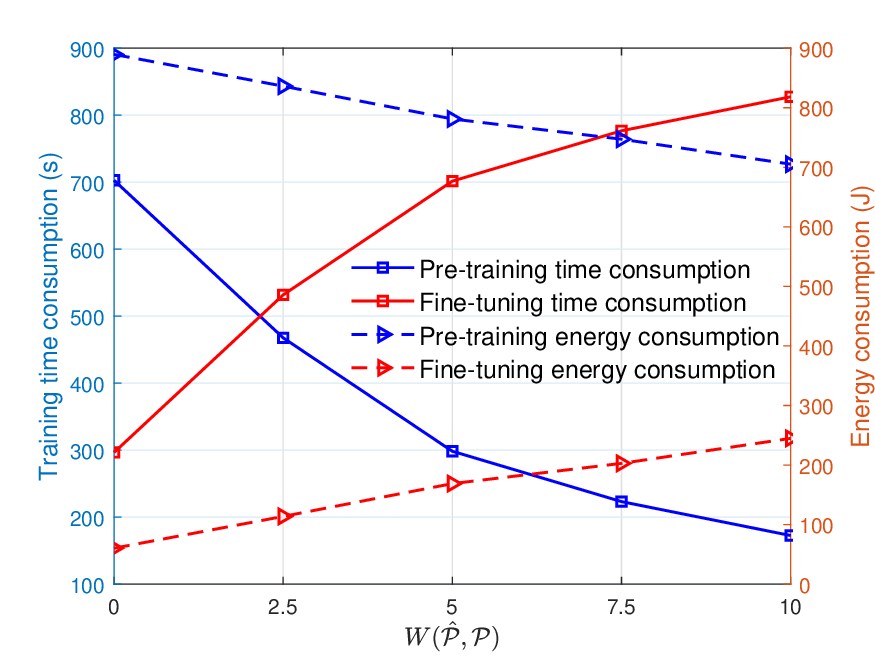}
	\vspace{-11pt}
	\caption{Total energy consumption and training delay of model pre-training and task-specific fine-tuning stages w.r.t. different data distribution distances $W(\hat{\mathcal{P}},\mathcal{P})$.}
	\label{data_difference}
	\vspace{-14pt}
\end{figure}

Fig. \ref{data_difference} shows the total energy consumption and training delay of model pre-training and task-specific fine-tuning stages w.r.t.  varying data distribution distances $W(\hat{\mathcal{P}},\mathcal{P})$ in \eqref{convergence analysis}. It is observed that, as the data distribution between the pre-training and fine-tuning tasks gets more diverse (i.e., $W(\hat{\mathcal{P}},\mathcal{P})$ increases), the system energy consumption and training delay decrease for the pre-training stage, while those for the fine-tuning stage increase. This is because, as the datasets in model pre-training and fine-tuning stages get more diverse, we generally should allocate more resources on the task-specific fine-tuning stage to better align with the target task for enhancing the ultimate learning performance.

\vspace{-6pt}
\section{Conclusion}
This paper has studied the entire life-cycle of edge learning including two stages of model pre-training and task-specific fine-tuning. Specifically, model pre-training was conducted via centralized learning, and task-specific fine-tuning was performed via FEEL. For the proposed  system, we first theoretically analyzed the average squared gradient norm bound, overall energy consumption, and learning delay. Then, based on the analytical results, we optimized the computation and communication resources (including learning rounds, batch sizes,  clock frequencies, and transmit power of both pre-training and fine-tuning stages) to minimize the average gradient bound, subject to the transmit power, energy consumption, and delay constraints. Next, we proposed an efficient algorithm to solve the highly non-convex average squared gradient bound minimization problem. Finally, numerical results validated that the proposed design  achieved better learning performance than benchmark schemes. How to extend the theoretical analysis in scenarios with modified  model architectures across the pre-training and fine-tuning stages, and extend to other scenarios (with, e.g., AirComp and renewable energy supplies) are interesting topics worth pursuing in future work.

\vspace{-5pt}
\begin{appendix}
\subsection{Proof of Lemma \ref{lemma1}}
To start with, we bound the improvement of the loss for a single step. Based on \eqref{FEEL parameter update} and Assumption 2, we have
\begin{align} \label{imrove finetune}
	&{L}(\hat{\mv w}^{(n+1)},{\cal B}) - {L}(\hat{\mv w}^{(n)},{\cal B}) \nonumber \\
	&\le {\big(\nabla L(\hat{\mv w}^{(n)},{\cal B})\big)^{\rm T}}(\hat{\mv w}^{(n+1)} - \hat{\mv w}^{(n)}) + \frac{\hat \rho}{2}\|\hat{\mv w}^{(n+1)} -\hat{\mv w}^{(n)}\|^2 \nonumber \\
	& \le- {\big(\nabla {L}(\hat{\mv w}^{(n)},{\cal B})\big)^{\rm T}}\left( \hat{\gamma} \sum\limits_{k \in {{\mathcal K}}} \frac{\tilde{B}^{(n)}_{k}}{\tilde{B}^{(n)}_{\rm tot}} \nabla \tilde{{{ L}}}^{(n)}_{k} (\hat{{\mv w}}^{(n)}, \tilde{\mathcal B}^{(n)}_{k}) \right) \nonumber \\
	&\qquad + \frac{{{\hat \rho} \hat{\gamma}^2 }}{2} \left\| \sum\limits_{k \in {{\mathcal K}}} \frac{\tilde{B}^{(n)}_{k}}{\tilde{B}^{(n)}_{\rm tot}} \nabla \tilde{{{ L}}}^{(n)}_{k} (\hat{{\mv w}}^{(n)}, \tilde{\mathcal B}^{(n)}_{k}) \right\|^2.
\end{align}

Then, by taking the expectation at both sides of \eqref{imrove finetune}, and it follows that 
\begin{align} \label{improve finetune expectation}
	&{\mathbb E}\left[{L}(\hat{\mv w}^{(n+1)},{\cal B})\right] - {\mathbb E}\left[{L}(\hat{\mv w}^{(n)},{\cal B}) \right] \nonumber \\
	& \le - \hat\gamma   {\mathbb E} \left[\sum\limits_{k \in {{\mathcal K}}} \frac{\tilde{B}^{(n)}_{k}}{\tilde{B}^{(n)}_{\rm tot}} \big(\nabla {L}(\hat{\mv w}^{(n)},{\cal B})\big)^{\rm T} \nabla \tilde{{{ L}}}^{(n)}_{k} (\hat{{\mv w}}^{(n)}, \tilde{\mathcal B}^{(n)}_{k})\right] + \nonumber \\
	& \qquad \frac{{{\hat \rho} \hat{\gamma}^2 }}{2}  {\mathbb E} \left[ \left\| \sum\limits_{k \in {{\mathcal K}}} \frac{\tilde{B}^{(n)}_{k}}{\tilde{B}^{(n)}_{\rm tot}} \nabla \tilde{{{ L}}}^{(n)}_{k} (\hat{{\mv w}}^{(n)}, \tilde{\mathcal B}^{(n)}_{k}) \right\|^2\right] \nonumber \\
	& \le \!\!-\! \hat{\gamma} {\mathbb E} \!\! \left[\! \| \! \nabla\! {L}(\hat{\mv w}^{(n)}\!\!\!,{\cal B}) \! \|^2 \! \right] \!\!+\!\! \frac{{{\hat \rho} \hat{\gamma}^2 }}{2} \!\! \left( \!\!{\mathbb E}\!\! \left[ \!\| \!\nabla \!{ L}(\hat{\mv w}^{(n)}\!\!\!,{\cal B}) \! \|^2\! \right] \!\!+\!\! \frac{K \hat{\alpha}^2}{\tilde{B}^{(n)}_{\rm tot}} \!\!\right)\!,\!
\end{align}
where the last inequality is obtained from Assumption 3 and 
\begin{align} 
&\mathbb{E}\left[ \left\| \sum\limits_{k \in {{\mathcal K}}} \frac{\tilde{B}^{(n)}_{k}}{\tilde{B}^{(n)}_{\rm tot}} \nabla \tilde{{ L}}^{(n)}_{k} (\hat{{\mv w}}^{(n)}, \tilde{\mathcal B}^{(n)}_{k}) \right\|^2  \right] \nonumber \\
&\le \sum\limits_{k \in {{\mathcal K}}} \frac{\tilde{B}^{(n)}_{k}}{\tilde{B}^{(n)}_{\rm tot}}\mathbb{E}\left[ \| \nabla \tilde{{ L}}^{(n)}_{k} (\hat{{\mv w}}^{(n)}, \tilde{\mathcal B}^{(n)}_{k}) \|^2 \right] \nonumber \\
& \le \| \nabla {L}(\hat{\mv w}^{(n)},{\cal B}) \|^2+\frac{K \hat{\alpha}^2}{\tilde{B}^{(n)}_{\rm tot}},
\end{align}
where the first and second inequalities follow from the Jensen's inequality and Assumption 3, respectively. Finally, by combining  both sides of \eqref{improve finetune expectation} for $n \in {\mathcal N}$, Lemma \ref{lemma1} is proved.

\subsection{Proof of Lemma \ref{lemma2}}
In the proof, we first analyze the total expected loss reduction in the pre-training stage, and then characterize the change of loss when shifting data distributions to the fine-tuning task. In such a way, we could characterize the contribution of the pre-trained model ${\mv w}^{(M)}$ as the initial model $\hat{\mv w}^{(0)}$  for task-specific fine-tuning in terms of expected loss reduction. To start with, for single pre-training step $m$, based on \eqref{parameter update pre-train} and Assumption 2, we have
\begin{align} \label{improve pretrain}
&L({\mv w}^{(m + 1)},{\cal D}) - L({\mv w}^{(m )},{\cal D}) \nonumber \\
&\le - {\big(\nabla L({\mv w}^{(m)},{\cal D})\big)^{\rm T}}\left(\gamma\nabla \tilde L({\mv w}^{(m)},\tilde {\mathcal D}^{(m)})\right) \nonumber \\
&\qquad + \frac{{\rho \gamma^2 }}{2}\|\nabla \tilde L({\mv w}^{(m)},\tilde {\mathcal D}^{(m)})\|^2.
\end{align}

Next, by taking the expectation on both sides of \eqref{improve pretrain}, we have 
\begin{align} \label{improve pretrain expectation}
	&{\mathbb E} \!\left[\!L({\mv w}^{(m+1)}\!,{\cal D})\!\right] \!- \!{\mathbb E}\!\left[L({\mv w}^{(m)},{\cal D})\!\right]  \!\le\! \!- \gamma {\mathbb E} \!\left[ \!\| \nabla L({\mv w}^{(m)}\!,{\cal D}) \|^2 \!\right] \nonumber  \\
	& \qquad + \frac{{\rho \gamma^2 }}{2} \!\left( \!{\mathbb E} \left[ \| \nabla L({\mv w}^{(m)},{\cal D}) \|^2 \right] \!\!+\!\! \frac{\alpha ^2} {\tilde{D}^{(m)}}\right)\!,
\end{align}
where the last inequality follows from Assumption 3. Finally, combining  both sides of \eqref{improve pretrain expectation} for $m \in {\mathcal M}$, we obtain 
\begin{align} \label{summed improve pretrain expectation}
	&{\mathbb E}\left[L({\mv w}^{(M)},{\cal D})\right] -{\mathbb E}\left[L({\mv w}^{(0)},{\cal D})\right] \le \nonumber \\
	& \!- \!\!\!\sum\limits_{m = 0}^{M-1}  \!\!\!\gamma  (1\!-\!\frac{{\rho \gamma }}{2} ) {\mathbb E} \!\left[\! \| \nabla L({\mv w}^{(m)}\!,{\cal D}) \|^2 \right] \!\!+\!\! \frac{{\rho \alpha^2 \gamma^2}}{2}  \!\!\sum\limits_{m = 0}^{M-1} \!\!\frac{1} {\tilde{D}^{(m)}}.
\end{align}

Then, we analyze the influence of changing datasets (from $\cal D$ with distribution  $\cal P$ to $ \cal B$ with distribution  $ \hat{\cal P}$) on the pre-trained model ${\mv w}^{(M)}$ in terms of the change of loss value. Specifically, with the upper bound on ${\mathbb E}\left[L({\mv w}^{(M)},\cal D)\right]$ on the pre-training dataset $\cal D$ in \eqref{summed improve pretrain expectation}, we have 
\begin{align} \label{changing data}
	&{\mathbb E}\left[L(\hat{\mv w}^{(0)},{\cal B})\right] - {\mathbb E}\left[L({\mv w}^{(M)},{\cal D})\right]  \nonumber \\
	&\le {\mathbb E}\left[| L(\hat{\mv w}^{(0)},{\cal B}) - L({\mv w}^{(M)},{\cal D})| \right] \nonumber \\ 
	& = \mathop {\sup }\limits_{{\mv u} \in {{\mathbb R}^q}:\|{\mv u}\| \le 1} \bigg\{ {\mathbb E}\left[ {{\mv u}^{\rm T}} \left(L(\hat{\mv w}^{(0)},{\cal B}) - L({\mv w}^{(M)},{\cal D})\right) \right] \bigg\} \nonumber \\
	&= \mathop {\sup }\limits_{{\mv u} \in {{\mathbb R}^q}:\|{\mv u}\| \le 1} \mathbb E \bigg\{  {\mathbb E_{{b_i} \sim \hat{\mathcal{P}}}}\left[{{\mv u}^{\rm{T}}}l(\hat{\mv w}^{(0)},b_i)\right] \nonumber\\ 
	& \qquad - {\mathbb E_{{d_{j}} \sim \mathcal{P}}}\left[{{\mv u}^{\rm{T}}}l({\mv w}^{(M)},d_{j})\right]  \bigg\} \le \tilde{\rho}W(\hat{\mathcal{P}},\mathcal{P}),
\end{align}
where the last inequality follows from the following lemma.
\begin{lemma}\label{Wasserstein lemma}
For any ${\mv u} \in {{\mathbb R}^q}$ with $\|{\mv u}\| \le 1$, we have
\begin{align}\label{Wasserstein lemma equation}
&{\mathbb E_{b_i \sim \hat{\mathcal{P}}}}\left[{{\mv u}^{\rm{T}}}l(\hat{\mv w}^{(0)},b_i)\right] - {\mathbb E_{{d_{j}} \sim \mathcal{P}}}\left[{{\mv u}^{\rm{T}}}l({\mv w}^{(M)},d_{j})\right] \nonumber \\
&\le \tilde{\rho}W(\hat{\mathcal{P}},\mathcal{P}).
\end{align}
\begin{IEEEproof}
	See Appendix C.
	\end{IEEEproof}
\end{lemma}

Finally, by substituting \eqref{summed improve pretrain expectation} into \eqref{changing data}, Lemma \ref{lemma2} is proved.

\subsection{Proof of Lemma \ref{Wasserstein lemma}}
To show that ${\mathbb E_{b_i \sim \hat{\mathcal{P}}}}\!\! \left[{{\mv u}^{\rm{T}}}l(\hat{\mv w}^{(0)}\!\!,b_i)\right] - {\mathbb E_{{d_{j}} \sim \mathcal{P}}}\!\big[{{\mv u}^{\rm{T}}}l({\mv w}^{(M)}\!\!,d_{j})\big]$ is upper bounded by $\tilde{\rho}W(\hat{\mathcal{P}},\mathcal{P})$, we first  show ${{\mv u}^{\rm{T}}}l({\mv w},d)$ is $\tilde{\rho}$-Lipschitz. This follows from 
\begin{align}
&\| {{\mv u}^{\rm{T}}}l({\mv w},b_i)-{{\mv u}^{\rm{T}}}l({\mv w},d_{j}) \| \nonumber \\
&\le \|{\mv u}\| \|l({\mv w},b_i)-l({\mv w},d_{j}) \| \le \tilde{\rho} dist(b_i,d_{j}),
\end{align}
where the first inequality follows from the Cauchy-Schwarz inequality and the second inequality is obtained from Assumption 4.

Next, we derive the desired upper bound. Assume $\mathcal{P}$ and $\hat {\mathcal{P}}$ have bounded support\footnote{It is worth noticing that such an assumption holds in general, as the data itself is always bounded or we normalize the data in practice.}. Then according to the Kantorovich-Rubinstein duality \cite{CVillani}, for a continuous function $f(d)$ with Lipschitz coefficient no larger than 1, the Wasserstein distance serves as a supremum as
\begin{align}\label{W_supremum}
W(\hat{\mathcal{P}},{\mathcal{P}})=\sup \bigg\{{{\mathbb E}_{b_i \sim \hat{\mathcal{P}}}}[f(b_i)] - {{\mathbb E}_{{d_{j}} \sim \mathcal{P}}}[f(d_{j})] \bigg\}.
\end{align}

Finally, by defining $f(d) \buildrel \Delta \over = \frac{1}{\tilde{\rho}}{{\mv u}^{\rm{T}}}l({\mv w},d)$, \eqref{Wasserstein lemma equation} is obtained from \eqref{W_supremum}. This thus completes the proof.
\end{appendix}

\end{document}